\documentclass[a4paper,aps,prd,twocolumn,eqsecnum,amsmath,amssymb,nofootinbib,superscriptaddress]{revtex4-2}
\usepackage{dcolumn}
\usepackage{bm}
\usepackage{graphics}
\usepackage{graphicx}
\usepackage{epsfig}
\usepackage{booktabs,multirow}
\usepackage{hhline}
\usepackage{slashed}
\usepackage{rccol}
\usepackage{mathrsfs}
\usepackage{amsmath, amssymb}
\usepackage[dvipsnames]{xcolor,colortbl}
\usepackage{subfigure}

\usepackage{xcolor}\colorlet{linkequation}{blue}
\usepackage[colorlinks=true, 
linkcolor=red,   
filecolor=red,   
citecolor=green, 
urlcolor=blue,
hyperfootnotes=true]{hyperref}

\newcommand*{\SavedEqref}{}
\let\SavedEqref\eqref
\renewcommand*{\eqref}[1]{%
	\begingroup
	\hypersetup{
		linkcolor=linkequation,
		linkbordercolor=linkequation,
	}%
	\SavedEqref{#1}%
	\endgroup
}

\begin{document}
\title{Solar neutrino constraints on light mediators through coherent elastic neutrino-nucleus scattering}

\author{M.~Demirci}
\email{mehmetdemirci@ktu.edu.tr}%
\affiliation{Department of Physics,
Karadeniz Technical University, Trabzon, TR61080, Türkiye}

\author{M.~F.~Mustamin}
\email{mfmustamin@ktu.edu.tr}
\affiliation{Department of Physics,
Karadeniz Technical University, Trabzon, TR61080, Türkiye}

\date{\today}

\begin{abstract}
We investigate new physics with light-neutral mediators through coherent elastic neutrino-nucleus scattering (CE$\nu$NS) at low energies. These mediators, with a mass of less than $1$ GeV, are common properties for extensions of the Standard Model (SM).
We consider general scalar, vector, and tensor interactions allowed by Lorentz invariance and involve universal light mediators accordingly.  In addition, we study an additional vector gauge boson with an associated $U(1)'$ gauge group for a variety of models including $U(1)_{B-L}$, $U(1)_{B-3L_e}$, $U(1)_{B-3L_\mu}$, and $U(1)_{B-3L_\tau}$. These models differ in the fermion charges, which determine their contributions within the CE$\nu$NS process. 
The effects of each model are investigated by embedding them in the SM process using solar neutrino flux. We derive new limits on the coupling-mass plane of these models from the latest CDEX-10 data. We also present projected sensitivities involving the future experimental developments for each model. Our results provide more stringent constraints in some regions, compared to previous works. Furthermore, the projected sensitivities yield an improvement of up to one order of magnitude.
\end{abstract}


\maketitle
\section{Introduction}
Coherent elastic neutrino-nucleus scattering (CE$\nu$NS) is a process in which neutrinos scatter off a nucleus that acts as a single particle \cite{Freedman:1973yd}. The difficulty in observing this process is due to its nuclear recoil energy that lies in the low keV range, which is the required criteria to ensure that neutrinos interact with the nucleus as a whole. This objective has finally been witnessed by the advancement of the COHERENT experiment \cite{Akimov:2017ade} on uncovering small recoil energies. In their first measurement, they used neutrinos emerging from pion decay at rest ($\pi$-DAR) at a Spallation Neutron Source (SNS) with a CsI[Na] scintillating crystal detector and then performed a second one using a liquid argon (LAr) detector \cite{Akimov:2020pdx}. Recently, they have updated the CsI data analysis with higher statistics along with an improved understanding of experiment systematics \cite{COHERENT:2021xmm}. This achievement has triggered scientific activities related to CE$\nu$NS, both experimentally and theoretically. This process provides a new promising framework to investigate fundamental parameters of the Standard Model (SM) and new physics beyond the SM (BSM). For instance, it has been widely used to study the weak mixing angle \cite{Cadeddu:2019,Cadeddu:2019eta}, the neutrino electromagnetic properties \cite{Khan:2023,Khan:2023b}, the effective generalized interactions \cite{Dent:2016wcr,  AristizabalSierra:2018eqm, Bischer:2019ttk, Flores:2021kzl}, the non-standard neutrino interactions (NSI) \cite{Barranco:2011wx,AristizabalSierra:2017joc,  Giunti:2019xpr, Mustamin:2021mtq}, dark-matter (DM) research \cite{Harnik:2012ni, Boehm:2020ltd,  Schwemberger:2022fjl}, and the light mediators \cite{Demirci:2021zci,  Coloma:2022umy,Majumdar:2021vdw, Cadeddu:2020nbr,DeRomeri:2022twg,AtzoriCorona:2022moj,Farzan:2018gtr,Li:2022jfl}.

One of the most intensive natural neutrino sources in the Earth is due to electron neutrinos produced by nuclear fusion processes inside the Sun. Solar neutrinos have been widely utilized since their first observation \cite{Davis:1968cp}. Solar neutrino measurements are tightly connected with the discovery of flavor conversion and the matter effect on neutrino dispersion which have been obtained thanks to the observation of different processes, such as charged-current (CC) \cite{Cleveland:1998nv, SAGE:1999uje}, neutral current (NC) \cite{SNO:2002tuh,  SNO:2008gqy}, and elastic scattering \cite{Kamiokande-II:1989hkh, Super-Kamiokande:2001ljr, Borexino:2007kvk, Borexino:2008fkj} that then lead to the formulation of the Standard Solar Model \cite{Bahcall:2000nu, Bahcall:2004mq, Bahcall:2004pz, Serenelli:2016dgz, Vinyoles:2016djt, Vitagliano:2019yzm}.  
Furthermore, the next generation of DM direct detection will be one of experiments that sensitive to coherent neutrino-nucleus scattering \cite{Cerdeno:2016sfi}.
The reason is that the CE$\nu$NS signal is closely related to DM direct-detection; both share a similar experimental signature: a sub-keV nuclear recoil and a similar detection technique based, e.g., on cryogenic bolometers.
DM direct detection experiments was proposed in the mid-eighties \cite{Drukier:1984vhf, Goodman:1984dc, Drukier:1986tm}. Recently,  such searches have already been carried out in various experiments such as LUX \cite{LUX:2015abn}, PandaX \cite{PandaX-II:2017hlx}, XENON \cite{XENON:2019gfn,XENON:2020kmp}, LZ \cite{LZ:2018qzl}, DarkSide \cite{DarkSide:2018kuk}, and, related to our main interest, CDEX \cite{CDEX:2013kpt, CDEX:2018lau, CDEX:2019isc, CDEX:2022mlp, Geng:2023yei}.
The synergy between CE$\nu$NS and DM searches expands to the characterisation of ``the neutrino floor'' \cite{OHare:2021utq}, which will ultimately limit the sensitivity of the current and next generation of experiments at low mass scales.
This limit depends on the input of particle, astrophysical, and nuclear models, and it is argued that it can be modified by the presence of non-standard neutrino-nucleus interactions. 

Although the SM presents a rather successful description of electroweak and strong interactions in nature, there are some shortcomings that point to the need to expand the current theory. In many extensions of the SM, low-mass particles appear from hidden sectors, such as grand unified theories \cite{Buchmuller:1991ce}, models that explain baryogenesis \cite{Fukugita:1986hr}, or dark sector models including a portal that provides a connection to SM particles \cite{Cirelli:2016rnw}. Many experiments around the world are now being developed to study light mediator models and hidden sectors (see, e.g., the review \cite{Essig:2013lka}). Experiments that are designed to detect CE$\nu$NS play a significant role in probing light mediator models. Since CE$\nu$NS process is well predicted in the SM, a measured deviation from it can provide a probing-ground of the BSM physics.

In regard to the aforementioned motivations, in this paper, we investigate the light mediator models beyond the SM through CE$\nu$NS with solar neutrinos. These models can be classified as follows:
\begin{itemize}
\item[i)] Universal light mediator models,
\item[ii)] $U(1)'$ models with $U(1)_{B-L}$ and $U(1)_{B-3L_\ell}$ $(\ell=e,\mu,\tau)$ gauge symmetries.
\end{itemize}

Among these, the universal light mediator models \cite{Cirelli:2013ufw, Abdallah:2015ter} are constructed to include only a few new particles and the possibility of Lorentz-invariant interactions. They can be considered as a limit of a more general new physics BSM scenario. Any other case can be derived from them by sufficient coupling rescaling.
In particular, we are going to focus on additional scalar, vector, and tensor mediators, which couple universally to all the SM fermions.  On the other hand, we also focus on an additional vector $Z'$ mediator with associated $U(1)'$ gauge group for a variety of models \cite{Allanach:2019} including $U(1)_{B-L}$ \cite{Mohapatra:1975, Mohapatra2:1975, Davidson:1978pm, Mohapatra:1980qe}, $U(1)_{B-3L_e}$, $U(1)_{B-3L_\mu}$, and $U(1)_{B-3L_\tau}$  \cite{Ma:1997nq, Ma:1998dr, Chang:2000xy, Heeck:2018nzc, Coloma:2020gfv}.
The vector $Z'$ mediator has an interaction term with all the SM quarks and leptons, and the couplings differ with the $U(1)'$ charges.
Accordingly, the CE$\nu$NS process can be utilized to study the vector boson $Z'$ by measuring deviations of the scattering cross section from the SM prediction.
All these proposals are theoretically well motivated ensuring a consistent description for a number of emerging discrepancies in precision studies of low-energy activities \cite{Muong-2:2006rrc,CDF:2022hxs}. These have also been phenomenologically studied through the CE$\nu$NS process with different neutrino sources such as $\pi$-DAR \cite{Cadeddu:2020nbr,DeRomeri:2022twg, AtzoriCorona:2022moj, Farzan:2018gtr}, reactors \cite{Billard:2018jnl, CONNIE:2019xid, CONUS:2021dwh}, and solar neutrino \cite{Li:2022jfl,Bertuzzo:2017} as well as next generation neutrino facilities \cite{Han:2019zkz, Bertuzzo:2022}.

We present new constraints on the coupling-mass parameters of light mediator models through  CE$\nu$NS using the recent CDEX-10 data \cite{CDEX:2022mlp}. The CDEX experiment, whose primary goal is to research light DM, has measured neutrino-nucleus event rates from solar neutrino flux using a p-type point contact germanium (PPCGe) detector array with $205.4$ kg$\cdot$day exposure. The CDEX-10 differential rate is given in terms of electron equivalent recoil energies that can be converted to nuclear recoil signals using a quenching factor. Moreover, the experiment will be upgraded to the other phase with a 50 kg germanium detector array, which is called CDEX-50 \cite{Geng:2023yei}. Accordingly, we also include the projected sensitivities to this improvement on the considered light mediator models. Furthermore, we compare our results with the existing limits of previous works, derived from BOREXINO \cite{Borexino:2007kvk}, COHERENT\cite{Akimov:2017ade, Akimov:2020pdx}, CONNIE \cite{CONNIE:2019xid}, CONUS \cite{CONUS:2021dwh}, TEXONO \cite{TEXONO:2009knm}, XENON \cite{XENON:2019gfn,XENON:2020kmp}, etc. 

We structure the remainder of our work as follows. In Sec.~\ref{sec:cevns}, we present the theoretical formulation of CE$\nu$NS in both the SM and the light mediator models.
In Sec.~\ref{sec:stat}, we lay out the data analysis method employed for limit setting.
In Sec.~\ref{sec:resdis}, for each model, we present the expected event spectra. Then we show new upper-limits on the allowed parameter space and compare them with the other current limits. Finally, we conclude our work in Sec.~\ref{sec:summ}.

\section{General Framework}\label{sec:cevns}
In this section, we first review the necessary details on cross-sections for the SM and the light mediator models beyond the SM, respectively. Then we give theoretical details on the event rates using solar neutrino flux. We also introduce the nuclear recoil energy conversion to its electron equivalent using the quenching factor.

\subsection{Standard formalism of CE$\nu$NS}
In the framework of the SM, the CE$\nu$NS process is well predicted. In this process, a neutrino with initial energy $E_\nu$ scatters from a nucleus target and imparts a kinetic recoil energy $T_{nr}$ to the nucleus. Coherent scattering occurs as a purely quantum effect where the initial neutrino has small enough energy, such that it is unable to probe the interior nucleon structure of the target. Particularly, for the transfer momentum $|\vec{q}| \lesssim \frac{1}{R}$ with the typical nuclear size $R$, the coherent scattering will provide an enhancement in the cross section.

\begin{figure}[!h]
	\centering
	\includegraphics[scale=0.9]{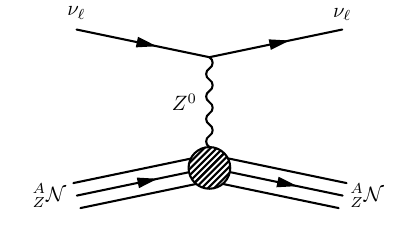}
	\caption{Feynman diagram for CE$\nu$NS in the SM. $^A_Z \mathcal{N}$ represents a nucleus with $A$ nucleons ($Z$ protons and $N=A-Z$ neutrons) and $Z^0$ is the SM neutral vector boson. The subscript $\ell$ denotes the neutrino flavour $\ell=e,\mu$ or $\tau$.}
	\label{fig:cevns}
\end{figure}

The Feynman diagram for CE$\nu$NS in the SM is shown in Fig.~\ref{fig:cevns}. The differential cross-section for CE$\nu$NS in the SM with respect to the nuclear recoil energy $T_{nr}$ is given by
\begin{align}
	\biggl[\frac{d\sigma}{dT_{nr}}\biggr]_{\text{SM}} =
	\frac{G_F^2 m_\mathcal{N}}{\pi} Q_{\text{SM}}^2 \left(1 - \frac{m_\mathcal{N} T_{nr}}{2E_\nu^2} \right) \bigl|F(|\vec{q}|^2)\bigr|^2, 
	\label{eq:cevns_difcross}
\end{align}
where $G_F$ is the Fermi constant and $m_\mathcal{N}$ is the nucleus mass. The weak nuclear charge $Q_{\text{SM}}$ is defined by
\begin{align}
	Q_{\text{SM}} = g_V^p Z+ g_V^n N,
	\label{eq:weak_chrg}
\end{align}
where the proton and neutron couplings are  $g_V^p=1/2(1-4\sin^2\theta_W)$ and $g_V^n=-1/2$, respectively. Although the proton contribution is small, it has a dependence on the fundamental weak-mixing angle.
For the weak-mixing angle, we use the value of $\sin^2\theta_W = 0.23863$ \cite{ParticleDataGroup:2022pth} obtained at low momentum transfer in the $\overline{\text{MS}}$ scheme.
We note that the expression \eqref{eq:cevns_difcross} is valid for both neutrinos and antineutrinos. Furthermore, the SM cross section of CE$\nu$NS is flavor-blind at the tree level, with small loop corrections that are flavor dependent but have no significant impact on current experimental sensitivities \cite{Tomalak:2021}.

The function $F(|\vec{q}|^2)$ is the weak nuclear form factor, which describes the nucleon complex structure of the target nucleus. Here we consider the same form factor for both proton and neutron, i.e., $F_p\simeq F_n=F$, and use the Helm parametrization, given by \cite{Helm:1956zz}
\begin{align}
F(|\vec{q}|^2) = 3 \frac{j_1(|\vec{q}|R_0) }{ |\vec{q}|R_0} e^{-\frac{1}{2}|\vec{q}|^2 s^2}
\label{eq:ffhelm}
\end{align}
where $j_1(x)=\sin x/x^2-\cos x/x$ is the  first order Spherical Bessel function and the three-momentum transfer is $|\vec{q}| = \sqrt{2m_\mathcal{N} T_{nr}}/197.3 $ fm$^{-1}$. The diffraction radius is defined by $R_0^2=\frac{5}{3}R^2-5s^2$, with  the nuclear radius $R=1.23A^{1/3}$  and the surface thickness $s=0.9$ fm.

\subsection{Light Mediator Models}
\subsubsection{Universal Light Mediators}
The universal light mediator models \cite{Cirelli:2013ufw, Abdallah:2015ter, Cerdeno:2016sfi} are constructed to include only a few new particles and interactions so that they can be considered as a limit of a more general new physics BSM scenario. By construction, this model can be characterized in terms of only a few free parameters such as coupling constants and masses.
Such models allow us to research broad classes of new physics signals without specifying any full high-energy theory.
We note that by making the new mediators heavy enough, it is possible to move from this model to the effective field theory framework.

In our endeavor, we consider general scalar, vector, and tensor interactions allowed by Lorentz invariance and involving light mediators. Such neutrino generalized interactions have been previously studied for CE$\nu$NS in Refs. \cite{Demirci:2021zci, Majumdar:2021vdw,Cadeddu:2020nbr, DeRomeri:2022twg, AtzoriCorona:2022moj, Farzan:2018gtr, Li:2022jfl}.
At low recoils, CDEX-10 data may place competitive or stronger bounds on their couplings and masses, as the new physics
effects are inversely proportional to the recoil kinetic energy. 

\vspace{0.2cm}
\textbf{Light scalar mediator}: We consider a new scalar mediator $\phi$, which mediates an interaction between neutrinos and quarks. The generic interaction Lagrangian can be written as \cite{Cerdeno:2016sfi}
\begin{align}
\begin{split}
\mathcal{L}_\phi = -  \phi &\biggl[\sum_q g_{\phi}^q \bar{q} q + g_{\phi}^{\nu_\ell} \bar{\nu_\ell}_R {\nu_\ell}_L  + h.c. \biggr] ,
\end{split}
\end{align}
where the scalar coupling constant $g_{\phi}^q$ is for q-quark and $g_{\phi}^{\nu_\ell}$ for neutrino. The label $\ell$ refers to lepton flavor $e, \mu,$ and $\tau$.
In this model, the contribution to the CE$\nu$NS process is obtained by 
\begin{align}
\biggl[\frac{d\sigma}{dT_{nr}}\biggr]_\phi = 
\frac{Q_\phi^2 T_{nr} m_\mathcal{N}^2 \bigl|F(|\vec{q}|^2)\bigr|^2}{4\pi(m_\phi^2+2m_\mathcal{N} T_{nr})^2 E_\nu^2}.
\label{eq:difcross_scalar}
\end{align}
The scalar charge of the nucleus $Q_\phi$ is defined as \cite{AristizabalSierra:2018eqm}
\begin{align}
&Q_\phi = \biggl[{Z}\sum_{q} g_\phi^{q} \frac{m_p}{m_q}f_{Tq}^p + {N}\sum_{q} g_\phi^{q} \frac{m_n}{m_q} f_{Tq}^n\biggr] g_\phi^{\nu_\ell},
\end{align}
where we set the hadronic structure parameters as $f_{T_u}^p = 0.0208$, $f_{T_u}^n = 0.0189$, $f_{T_d}^p = 0.0411$, and $f_{T_d}^n = 0.0451$ \cite{Hoferichter:2015tha}.
Notice that $m_p, m_n$, and $m_q$ are masses for proton, neutron, and quark, respectively. For simplicity, we consider equal couplings for the $u$ and $d$ quarks. This scalar interaction does not interfere with the SM. Hence, the new scalar mediator contribution adds incoherently to the SM CE$\nu$NS cross section.

\vspace{0.2cm}
\textbf{Light vector mediator}:
We consider a generic vector mediator $Z'$, which couples to both the SM quarks and neutrinos. The corresponding generic Lagrangian can be written as \cite{AtzoriCorona:2022moj} 
\begin{align}
\begin{split}
\mathcal{L}_{Z'} =Z'_\mu & \biggl[\sum_{q=u,d} Q'_{q} g_{Z'}^q \bar{q} \gamma^\mu q + Q'_{\ell} g_{Z'}^{\nu_\ell} \bar{\nu_\ell}_L \gamma^\mu  {\nu_\ell}_L  \biggr],
\end{split}
\end{align}
where $g_{Z'}^q$ and $g_{Z'}^{\nu}$ are vector coupling constants  for $q$-quark and  neutrino, respectively. The individual vector charges of quarks and neutrinos are given by $Q'_q$ and $Q'_\ell$ \footnote{We implement these charges for generalized form of anomaly free UV-complete models including only the SM particles plus right-handed neutrinos \cite{Allanach:2019}.}, respectively.
The vector mediator's contribution to the CE$\nu$NS cross section is given by
\begin{align}
\begin{split}
\biggl[\frac{d\sigma}{dT_{nr}}\biggr]_{Z'} &= 
\frac{Q_{Z'}^2 m_\mathcal{N} \bigl|F(|\vec{q}|^2)\bigr|^2}{2\pi(m_{Z'}^2 + 2m_\mathcal{N} T_{nr})^2} \left(1 - \frac{m_\mathcal{N} T_{nr}}{2E_\nu^2} \right), 
\end{split}
\label{eq:difcross_vector}
\end{align}
where $Q_{Z'}$ is the weak vector charge of the nucleus. Vector current conservation implies that only valence quarks contribute by simply adding their charges,
so we have
\begin{align}
\begin{split}
Q_{Z'} &= \biggl[{Z} \sum_q Q'_{q} g_{Z'}^q + {N} \sum_q Q'_{q} g_{Z'}^q \biggr] Q'_{\ell} g_{Z'}^{\nu_\ell} \\
&= \biggl[{Z} (2Q'_{u} + Q'_{d}) + {N}  (Q'_{u} + 2Q'_{d}) \biggr] g_{Z'}^{q} g_{Z'}^{\nu_\ell} Q'_{\ell}.
\end{split}
\label{eq:zpnuccharge}
\end{align}

We give the above relation on $Q_{Z'}$ in the general form in terms of charges $Q'_{u}$, $Q'_{d}$, and $Q'_{\ell}$. However, in the framework of the universal mediator, the vector mediator couples universally to all the SM fermions. Accordingly, we set $Q'_{u}=Q'_{d}=Q'_{\ell}=1$ in Eq. \eqref{eq:zpnuccharge} for this model.

Since both the SM and the $Z'$ interactions are of vector type, they contribute coherently to the CE$\nu$NS cross section. The $Z'$ mediator has an interference term with the SM case, so we have the complete cross section as follows:
\begin{align}
\begin{split}
\biggl[\frac{d\sigma}{dT_{nr}}\biggr]_{\text{SM}+Z'} =& \left[1+\frac{Q_{Z'}}{\sqrt{2} G_F Q_{\text{SM}} (m_{Z'}^2+2m_\mathcal{N} T_{nr})}\right]^2 \\ & \times \biggl[\frac{d\sigma}{dT_{nr}}\biggr]_{\text{SM}}.
\end{split}
\label{eq:difcross_vectorint}
\end{align}

\vspace{0.2cm}
\textbf{Light tensor mediator}: We can write the interaction Lagrangian for a tensor $T$ mediator as
\begin{align}
\begin{split}
\mathcal{L}_{T} = &\biggl[\sum_q g_T^{q} \bar{q} \sigma^{\mu\nu} q -  g_T^{\nu_l} \bar{\nu_\ell}_R \sigma^{\mu\nu} {\nu_\ell}_L\biggr] T_{\mu\nu} ,
\end{split}
\end{align}
where $\sigma_{\mu\nu} =
i(\gamma_\mu\gamma_\nu-\gamma_\nu\gamma_\mu)/2$. The $g_{T}^q$ and  $g_{T}^{\nu_\ell}$ are tensor coupling constants for quarks and neutrinos, respectively.  Note that this mediator does not interfere with the SM case. The relevant contribution to the CE$\nu$NS process is given by \cite{DeRomeri:2022twg}
\begin{align}
\begin{split}
\biggl[\frac{d\sigma}{dT_{nr}}\biggr]_{T} &= 
\frac{2 Q_{T}^2 m_\mathcal{N} \bigl|F(|\vec{q}|^2)\bigr|^2}{\pi (m_{T}^2 + 2m_\mathcal{N} T_{nr})^2}  \left(1 - \frac{m_\mathcal{N} T_{nr}}{4E_\nu^2} \right), 
\end{split}
\label{eq:difcross_tensor}
\end{align}
with
\begin{align}
&Q_T = \biggl[{Z}\sum_{q} g_T^{q} \delta_q^p + {N}\sum_{q} g_T^{q} \delta_q^n \biggr] g_T^{\nu_l}.
\end{align}
We set the parameters as $\delta^p_u=\delta^n_d = 0.54$, and $\delta^p_d=\delta^n_u = -0.23$ \cite{Cirelli:2013ufw}. There is also a different set of delta parameters (see, Ref. \cite{Cirelli:2013ufw}). We make our choice for consistency with recent works \cite{DeRomeri:2022twg,Majumdar:2021vdw}.

\subsubsection{Vector mediator from $U(1)'$ models}
There are many studies of SM extensions with the addition of a $U(1)'$ gauge group with an associated neutral gauge boson $Z'$ (see, e.g., Ref. \cite{Langacker:2009} for review). A necessary condition is that the theory is anomaly-free. Anomaly-free models can be constructed by expanding the SM with three right-handed neutrinos. Such expansion simultaneously explains smallness of neutrino mass through the see-saw mechanism \cite{Okada:2018ktp}. Moreover, such models can explain some unsolved puzzles in the SM such as grand unified theory \cite{Buchmuller:1991ce}, baryogenesis mechanism through leptogenesis \cite{Fukugita:1986hr}, nature of DM \cite{Alves:2015pea}, and anomalies from experiments \cite{Allanach:2015gkd, Allanach:2023uxz}.

\begin{table}[h!]
	\caption{The $U(1)'$ charges of quarks and leptons for each model.}
	\begin{center}
	\begin{ruledtabular}
		\begin{tabular}{l  c  c r r r}
			Model & $Q'_{u}$& $Q'_{d}$ & $Q'_{e}$ & $Q'_{\mu}$ & $Q'_{\tau}$\\
			\hline
			universal & $1$ & $1$ & $1$ & $1$ & $1$ \\
			\hline
			$B-L$ & $1/3$ & $1/3$ & $-1$ & $-1$ & $-1$ \\
			$B-3L_e$ & $1/3$ & $1/3$ & $-3$ & $0$ & $0$ \\
			$B-3L_\mu$ & $1/3$ & $1/3$ & $0$ & $-3$ & $0$ \\
			$B-3L_\tau$ & $1/3$ & $1/3$ & $0$ & $0$ & $-3$ \\
		\end{tabular}
		\end{ruledtabular}
	\end{center}
	\label{tab:U1charge}
\end{table}

In this study, we focus on an additional vector $Z'$ mediator with an associated $U(1)'$ gauge group for a variety of models including $U(1)_{B-L}$ \cite{Mohapatra:1975, Mohapatra2:1975, Davidson:1978pm, Mohapatra:1980qe}, $U(1)_{B-3L_e}$, $U(1)_{B-3L_\mu}$, and $U(1)_{B-3L_\tau}$ \cite{Ma:1998dr, Chang:2000xy} (where $B$ stands for the baryon number and $L$ is for the lepton number). These models differ in terms of the charges of the fermions with associated gauge group. In Table \ref{tab:U1charge}, we list the $U(1)'$ charges of quarks and leptons for each model. It is seen that the vector $Z'$ mediator couples to the quarks and neutrinos with different charges. This difference determines the contributions of each model to CE$\nu$NS. Accordingly, the corresponding differential cross-sections are obtained from Eq.\eqref{eq:difcross_vectorint} by implementing these charges to Eq.\eqref{eq:zpnuccharge}. Hence, these contributions add coherently to the weak neutral current of the SM which is mediated by the $Z$ vector boson. The effects are quantified by additional terms in the nucleus weak charge.

Since these models depend on different neutrino flavors, we consider the solar neutrino survival probabilities, which arise due to neutrino propagation from the Sun to the Earth, resulting neutrino oscillations.  Accordingly, the differential cross sections are weighted by these survival probabilities. For the case of $\nu_e\rightarrow \nu_e$, we have
\begin{align}
P_{ee} = \cos^4{\theta_{13}} P_{eff} + \sin^4{\theta_{13}}.
\end{align}
While for the $\nu_e\rightarrow \nu_\mu$ and $\nu_e\rightarrow \nu_\tau$ the probabilities are
\begin{align}
P_{e\mu} = (1-P_{ee})\cos^2{\theta_{23}}, \\
P_{e\tau} = (1-P_{ee})\sin^2{\theta_{23}},
\end{align}
respectively. The factor $P_{eff}$ is the matter effect that satisfies \cite{ParticleDataGroup:2022pth}
\begin{align}
P_{eff} = \sin^2{\theta_{12}},
\end{align}
for solar neutrino in a few MeV energy. The probabilities are evaluated using the best-fit central values of the recent oscillation parameters with normal ordering  \cite{deSalas:2020pgw}.

\subsection{Differential Rate Spectra}
The event rate of the process, which is calculated by the convolution of cross section with neutrino flux, can be written as
\begin{align}
	\frac{dR}{dT_{nr}} = N_{T} \int_{E_{\nu}^{min}}^{E_{\nu}^{max}} dE_\nu \frac{d\Phi(E_\nu)}{dE_\nu} \frac{d\sigma(E_\nu, T_{nr})}{dT_{nr}},
\end{align}
where $d\Phi(E_\nu)/dE_\nu$ represents the differential flux of neutrinos.
The factor $N_{T}=m_t N_A/m_A$ denotes the number of target nuclei per unit mass of the detector material. Here, $m_t$ is the target mass, $m_A$ is the molar mass of the nuclei  and $N_A$ is the Avogadro's number.
The exposure of the CDEX-10 experiment is $205.4\text{ kg} \cdot \text{day}$ \cite{CDEX:2022mlp}.
The minimum and maximum  neutrino energy in the initial state are denoted by $E_{\nu}^{min}$ and $E_{\nu}^{max}$, respectively. The minimum neutrino energy satisfies
\begin{align}
	E_{\nu}^{min} = \frac{T_{nr}}{2}\left(1+\sqrt{1+\frac{2m_\mathcal{N}}{T_{nr}}} \right).
\end{align}
This energy is necessary to trigger the nuclear recoil energy. For the maximum neutrino energy, we use the endpoint of the solar neutrino flux spectrum. The maximum nuclear recoil energy obeys
\begin{align}
	T_{nr}^{max} = \frac{2E_\nu^2}{2E_\nu + m_\mathcal{N}}.
	\label{eq:Tnr_max}
\end{align}
This relation informs that lighter targets improve the maximum nuclear recoil energy produced in the detector.

The observed physical quantity in the experiment is different from the nuclear recoil energy signal as neutrinos scatter off the nuclei.
The detector observes an electron equivalent energy $T_{ee}$. Therefore, to relate these two quantities, the quenching factor $Y(T_{nr})$ is needed. For this purpose, we utilize two different quenching factors \footnote{We note that there are also other measurements \cite{Collar:2021fcl} regarding the quenching factor,  which could effect the derived limits.}. The first is the Lindhard quenching factor defined by \cite{Lindhard:1963}
\begin{align}
	Y(T_{nr}) = \frac{kg(\epsilon)}{1+kg(\epsilon)} 
	\label{eq:Linhard_quench}
\end{align}
with the parameters
\begin{align}
\begin{split}
	g(\epsilon) &= 3\epsilon^{0.15} + 0.7 \epsilon^{0.6} + \epsilon\\
	\epsilon =& 11.5 {Z}^{-7/3} T_{nr}. 
\end{split}
\end{align}
We note that the general form of $k$ is given by $k=0.133 {Z}^{2/3} {A}^{-1/2}$, but it is typically treated as a free-parameter since experimental results have various ranges of $k$ \cite{Schwemberger:2022fjl}. In this work, we set this parameter as $k=0.162$, which closely matches the recent measurement in the low-energy range \cite{Bonhomme:2022lcz}. The Linhard formula is acceptable for high recoil energy, namely $T_{nr}>0.254\text{ keV}$. Below this value, we use the another quenching factor, which is obtained from the ``high'' ionization-efficiency model \cite{Essig:2018tss}. For Ge target, this is given by
\begin{align}
Y(T_{nr}) = 0.18\left[1-e^{\left(\frac{15-T_{nr}}{71.03}\right)}\right],
\label{eq:lowquench}
\end{align}
which is acceptable in the range of $0.015 \text{ keV}<T_{nr}<0.254 \text{ keV}$. 

The nuclear recoil energy $T_{nr}$(keV) can be converted into electron equivalent energy $T_{ee}$(keV) with the quenching factor as
\begin{align}
	T_{ee} = Y(T_{nr}) T_{nr}.
\end{align}
From this form, we obtain the electron equivalency relation with the nuclear recoil energy as
\begin{align}
	\frac{dT_{ee}}{dT_{nr}} = Y(T_{nr}) + T_{nr} \frac{dY(T_{nr})}{dT_{nr}}.
\end{align}
Therefore, the differential rate in terms of the electron equivalent energy can be expressed as a function of the nuclear recoil energy:
\begin{align}
	\frac{dR}{dT_{ee}} = \frac{dR}{dT_{nr}} \frac{1}{Y(T_{nr})+T_{nr} \frac{dY(T_{nr})}{dT_{nr}}}.
\end{align}
In analyzing the CDEX-10 data which is in the range of $0.16 - 2.16$ keVee with a bin width of $100$ eVee, we use only the Linhard quenching in Eq.\eqref{eq:Linhard_quench}. This range is equivalent to $0.922 - 9.127$ keVnr, addressing the region of interest (ROI) of the experiment.

\begin{table}[h!]
	\caption{Solar neutrino fluxes with their uncertainties from the high-metallicity solar neutrino model BS05(OP). These values are taken from Refs. \cite{Bahcall:2004mq, Bahcall:2004pz}.}
	\begin{center}
	\begin{ruledtabular}
		\begin{tabular}{l c r }
			Components~ & Flux [$\text{cm}^{-2} \text{s}^{-1}$] & Uncertainty [$\%$] \\
			\hline
			pp & $5.99\times 10^{10}$ & $0.8\%$ \\
			pep & $1.42\times 10^{8}$ & $1.3\%$ \\
			hep & $7.93\times 10^{3}$ & $15.4\%$ \\
			$^8\mathrm{B}$ & $5.69\times 10^{6}$ & $12.6\%$ \\
			$^7\mathrm{Be}$ & $4.84\times 10^{9}$ & $9.3\%$ \\
			$^{13}\mathrm{N}$ & $3.07\times 10^{8}$ & $20.2\%$ \\
			$^{15}\mathrm{O}$ & $2.33\times 10^{8}$ & $23.3\%$ \\
			$^{17}\mathrm{F}$ & $5.84\times 10^{6}$ & $25.1\%$ \\
		\end{tabular}
		\end{ruledtabular}
	\end{center}
	\label{tab:solneu}
\end{table}

Neutrinos are produced inside the sun through a series of nuclear processes, which are commonly classified as the proton-proton (pp) chain or the Carbon-Nitrogen-Oxygen (CNO) cycle, depending on the elements involved.
Neutrinos of the pp chain come from five nuclear reactions referred to as the pp, pep, hep, $^8\mathrm{B}$, and $^7\mathrm{Be}$. In the CNO cycle, neutrinos are mainly produced from decays of $^{13}\mathrm{N}$, $^{15}\mathrm{O}$, and $^{17}\mathrm{F}$. In this work, we implement solar neutrino fluxes obtained from the high-metallicity solar neutrino model BS05(OP) \cite{Bahcall:2004mq, Bahcall:2004pz}.  In Table \ref{tab:solneu}, we list these fluxes together with their uncertainties.

\begin{figure*}[!htb]
	\centering
	\includegraphics[scale=0.45]{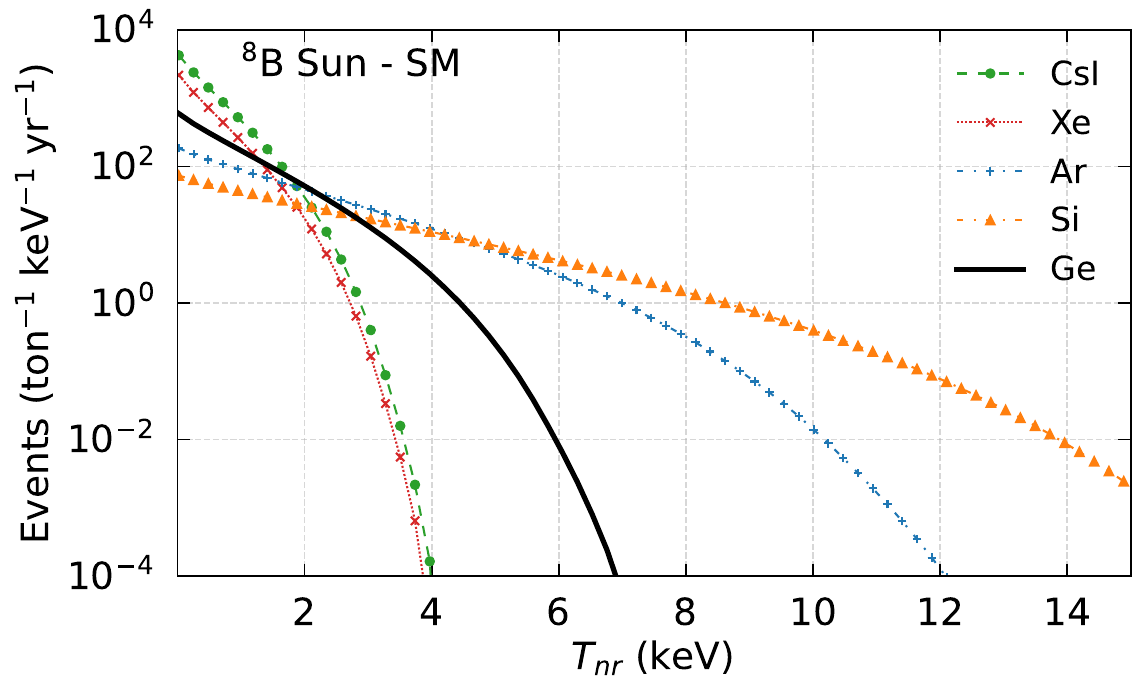}
	\includegraphics[scale=0.45]{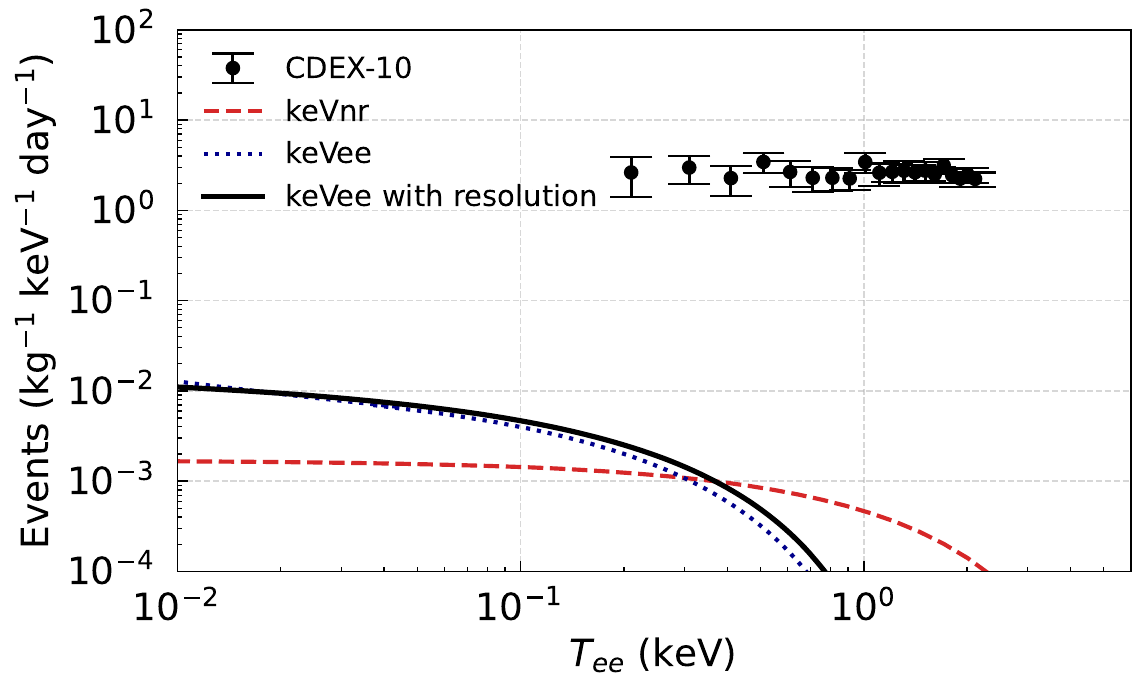}
	\\
 \hspace{10.4mm}	(a) \hspace{80.8mm} (b)
	\caption{(a) Event rate of the SM CE$\nu$NS process for several targets using $^{8}\mathrm{B}$ neutrino flux. (b) Effect of the quenching factor on the conversion between the nuclear recoil energy and its electron equivalency.
The red dashed line denotes the spectrum in keV in nuclear recoil energy. The blue dotted ones represent the electron equivalent conversion after implementing the quenching factor, namely for the high energy and low energy regions. The solid black line denotes the spectrum in keVee, implemented with recoil energy resolution.
	}
	\label{fig:cevnssm}
\end{figure*}

Integrating these neutrino fluxes over their energy spectra with the differential cross-section, we present the SM CE$\nu$NS event-rate with $^8\mathrm{B}$ neutrino flux in the unit of  $\text{ton}^{-1} \text{keV}^{-1} \text{year}^{-1}$ and $\text{kg}^{-1} \text{keV}^{-1} \text{day}^{-1}$
in Fig. \ref{fig:cevnssm}(a) and (b), respectively.
In Fig. \ref{fig:cevnssm}(a), we choose five different nucleus targets as follows: CsI, Xe, Ge, Ar, and Si. These are widely used as target in accelerator, reactor, and DM direct-detection experiments. Being the lightest nucleus, the CE$\nu$NS event rate for the Si target shows a relatively long-lived spectrum that reaches more than $14$ keV nuclear recoil energy. The event rates for CsI and Xe show a similar pattern since they have approximately the same atomic number.  The event rate for Ge target shows a wider spectrum than the ones of CsI and Xe. Relevant to our work is the case of the Ge nuclear target.

In Fig. \ref{fig:cevnssm}(b), we also present effect of the quenching factor on the conversion between the nuclear recoil energy and its electron equivalency. We examine the spectra of nuclear recoil energy (red dashed line), its electron equivalent energy (blue dotted line), and the effect of energy resolution $2.355 \times [35.8 + 16.6 \times T_{nr}(\text{ keV})]$. The CDEX-10 data  corresponds to the residual spectra with the $L$- and $M$-shell x-ray contributions subtracted in 0.16–2.16 keVee, at a bin width of 100 eVee \cite{CDEX:2022mlp}, which is shown by black dots with error bars. The event rate as a function of nuclear recoil energy (dashed-red) has a spectrum in order of $10^{-3}$ $\text{ kg}^{-1} \text{ keV}^{-1} \text{ day}^{-1}$ up to $2.3$ keVnr. The quenching factor effect increases the event rate (dotted-blue) to $10^{-2}$ $\text{ kg}^{-1} \text{ keV}^{-1} \text{ day}^{-1}$, while it ends at around $0.65$ keVee. The inclusion of the resolution function relatively preserves the shape of the event rate (solid-black), in which a bit of improvement can be seen in the high-energy region where it ends around $0.75$ keVee.

\begin{figure}[h]
	\centering
	\includegraphics[scale=0.43]{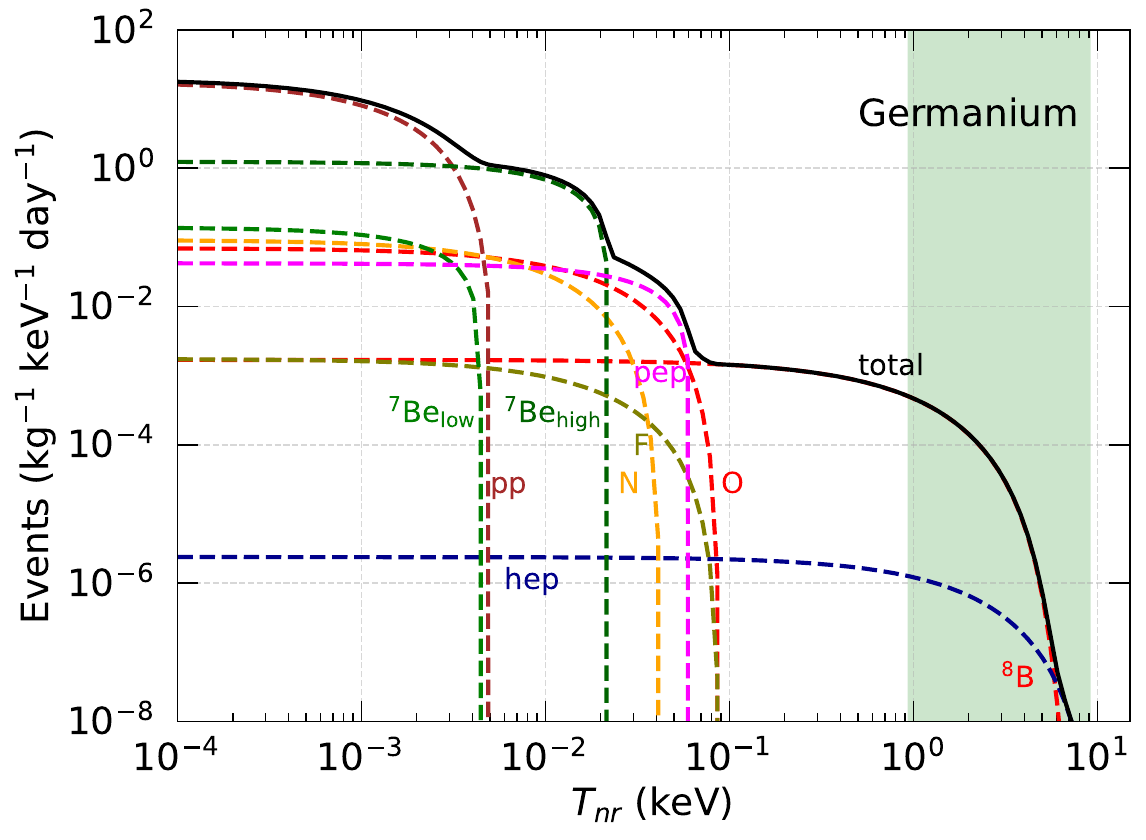}
	\caption{Expected spectra of CE$\nu$NS for Ge nuclear target as a function of nuclear recoil energy by using each solar neutrino flux (colored-dashed lines) and total flux (black-solid line). Here, we use the neutrino fluxes derived from the solar neutrino model BS05(OP) \cite{Bahcall:2004mq, Bahcall:2004pz}.
	}
	\label{fig:drate_total}
\end{figure}
In Fig. \ref{fig:drate_total}, we further show the differential event rate of CE$\nu$NS for the Ge target by using all the solar neutrino fluxes. The total event rate (solid-black) is obtained by summing up the contribution of neutrinos originated from the $pp$-chain and CNO-cycle (dashed-colored lines). Event rates in the high energy nuclear recoil region ($\gtrsim 0.1 \text{ keV}$)  are dominated by the hep and $^{8}\mathrm{B}$ neutrinos, while the others dominate in the small energy.  The shaded region indicates the ROI for the CDEX-10 result.

\section{Data Analysis Details}\label{sec:stat}
In this section, we describe the procedure employed for limit setting. In this study, we are interested in the CE$\nu$NS process with solar neutrinos in the CDEX-10 experiment \cite{CDEX:2013kpt,CDEX:2018lau, CDEX:2019isc, CDEX:2022mlp}. In the recent CDEX-10 work \cite{CDEX:2022mlp}, they present the event-rates of coherent neutrino-nucleus scattering, together with neutrino-electron. The CDEX experiment, which is part of the China Jinping Underground Laboratory (CJPL) \cite{CDEX:2013kpt}, has been dedicated to the direct detection of DM, using ultra-low energy threshold pPCGe detectors. Since first started, the experiment has been conducting some exotic physics searches such as WIMP, axion particle,  and dark photon. The CDEX-10 experimental configuration
has been described in Ref. \cite{CDEX:2018lau}. Both CE$\nu$NS and neutrino-electron scattering processes can enhance the observation limits of the light mediator models in the ROI of DM direct detection.
In this regard, we use the recent CDEX-10 data (20 data points) related to neutrino-nucleus scattering. These data are given in terms of electron-equivalent recoil energy. We convert this into the nuclear recoil energy by help of the quenching factors (see, for details, the previous section).

We adopt the pull approach of the $\chi^2$ function \cite{Fogli:2002pt}
\begin{widetext}
	\begin{align}
		\chi^2 = \mathrm{min}_{(\xi_j)} \sum_{i=1}^{20} &\Bigg( \frac{ R_{obs}^{i} - R_{exp}^{i} - B - \sum_j \xi_j c_{j}^i }{\Delta^{i}}\Bigg)^2  +  \sum_j \xi_j^2
	\end{align}
\end{widetext}
for constraining the investigated model parameters. Here, $R_{obs}^i$ and $R_{exp}^i$ are the observed and expected event rates (that consist of SM plus new physics contribution) respectively, in the $i$-th energy bin.
We note that the efficiency effect has been considered in calculating $R_{exp}^i$, which comes from the combination of the trigger efficiency and the physics-noise (PN) cut efficiency \cite{CDEX:2018lau}. The experimental uncertainty for the $i$-th energy bin is denoted by $\Delta^i$, which includes statistical and systematic uncertainties. The latter uncertainty mainly comes from the choices of background, sources, bin size, as well as the rise-time shift, which are listed in Table I of Ref.  \cite{CDEX:2018lau}.
Moreover, the solar neutrino flux uncertainty is represented by $c_j^i$. The function is minimized with respect to pull parameters $\xi_j$ \cite{Fogli:2002pt}.

One of our goals is to compare and contrast the results obtained from an ideal vs. realistic detector. DM direct detection experiments are in advancement and will be sensitive to the  CE$\nu$NS process with solar neutrinos, providing the possibility to catch clues to new physics beyond the SM at the low energy range. Development of the current CDEX experiment is under construction to enter this era.
The third phase of the experiment which is called CDEX-50 \cite{Geng:2023yei} plans to have $50$ kg of high purity germanium as its detector array. It aims to reduce the background to about $0.01 \text{ events } \text{keV}^{-1}\text{kg}^{-1} \text{day}^{-1}$ \cite{CDEX:2022mlp}. Exposure of this upgrade will be $50 \text{ kg}\cdot\text{year}$ and threshold of $160 \text{eVee}$ \cite{Geng:2023yei}. These future configurations are expected to further constrain the new physics parameters. Regarding this scheme, we propose two scenarios in this work. The first is a realistic scenario, in which we assume the experiment uncertainty could reach $10\%$. The second is an optimistic scenario, where the uncertainty is set to be $1.5\%$. Such small uncertainties are expected by the improved purity of detector components, stringently controlled germanium exposure, and tank shielding in the future experiment \cite{Geng:2023yei}. A flat background according to this advancement is assumed in the two considerations.
These scenarios are implemented in the ROI of the future observed recoil energy. The nuclear recoil would then be in the region of $0.1$ keV and $0.015$ keV for the considered realistic and optimistic scenarios, respectively. 

\begin{figure*}[!htb]
	\centering
	\includegraphics[scale=0.41]{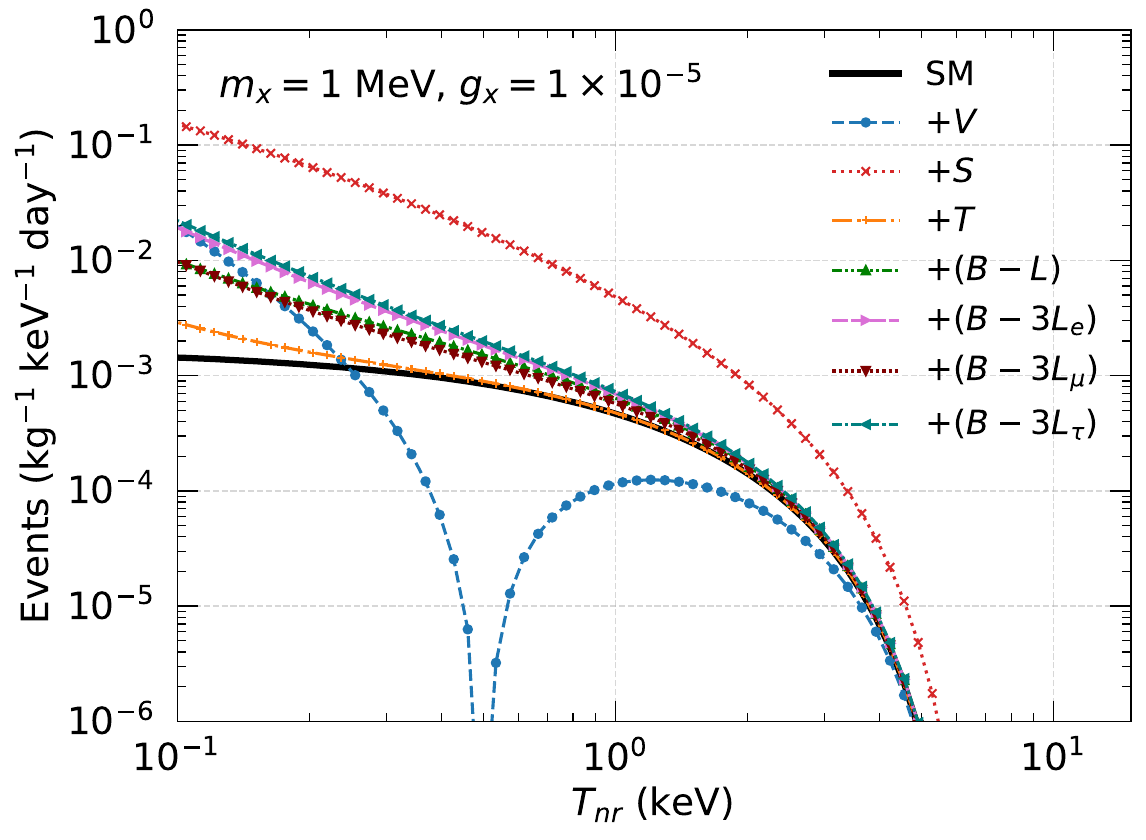}
	\includegraphics[scale=0.41]{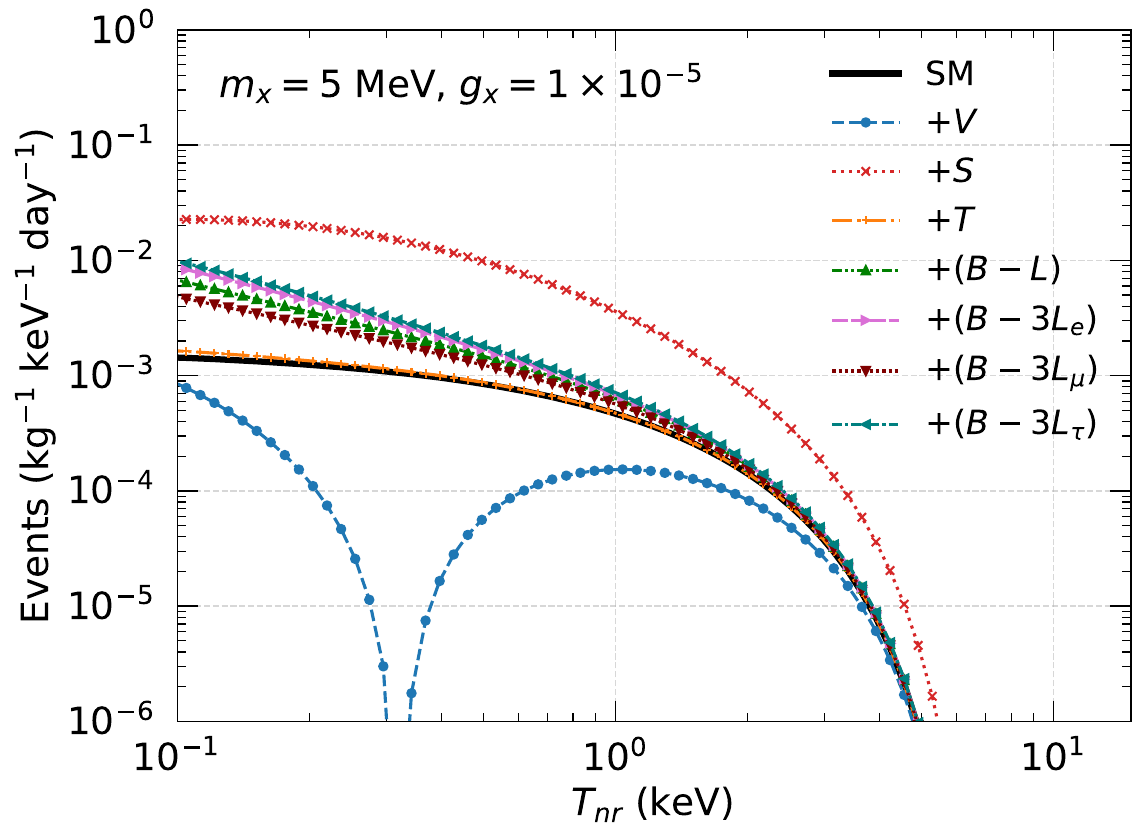}
	\\
 \hspace{10.4mm}	(a) \hspace{80.8mm} (b)
	\\
	\includegraphics[scale=0.41]{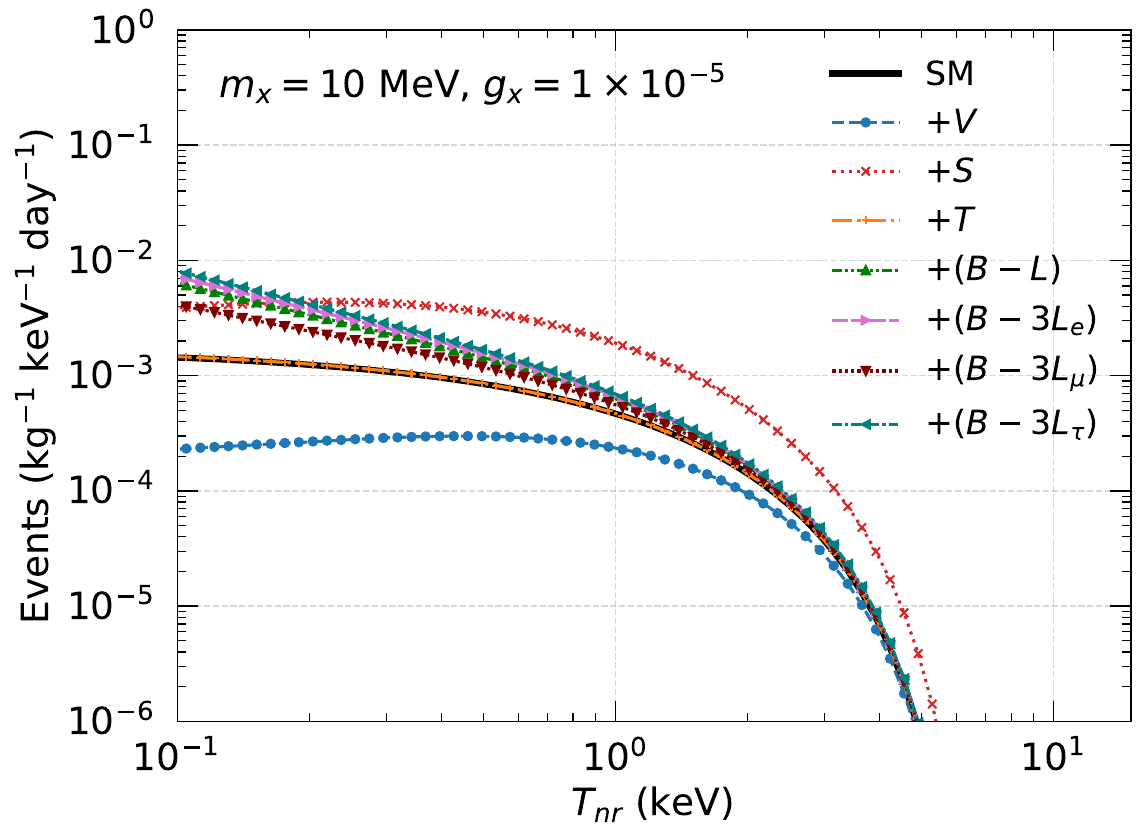}
	\\
	\hspace{10.4mm} (c)
	\caption{Predicted CE$\nu$NS differential rates as a function of nuclear recoil energy in the SM (solid black) and the light mediator models with a mass of (a) 1 MeV, (b) 5 MeV, and (c) 10 (MeV). Here, we use the $^8\mathrm{B}$ neutrino flux which has a large energy spectrum on the Earth.}
	\label{fig:cevns_bsm}
\end{figure*}
\section{Results and Discussion}\label{sec:resdis}
In this section, we provide numerical results from statistical analysis. First, we present the expected event spectra as a function of the nuclear recoil energy. Second, we present new constraints on the mass-coupling parameter space of the flavor-universal scalar, vector, and tensor mediator models, as well as the $U(1)'$ models. We also illustrate projected sensitivities for realistic and optimistic scenarios. We compare our results with the existing limits derived from further experimental probes.
\subsection{Expected Event Spectra}
The CE$\nu$NS differential event rates for the SM and the light mediator models are shown in Fig. \ref{fig:cevns_bsm}. They are normalized in $\text{ kg}^{-1} \text{ keV}^{-1} \text{ day}^{-1}$. We define the coupling constants as $g_X = \sqrt{g_X^{q}g_X^{\nu_l}}$, where the label $X = \phi, Z', T$, $B-L$, $B-3L_e$, $B-3L_\mu$, and $B-3L_\tau$ stand for the universal scalar, vector, tensor, and the corresponding $U(1)'$ mediators, respectively. For illustrative purposes, we set the coupling constant as $g_X=10^{-5}$ and the mediator mass as $1$ MeV in Fig. \ref{fig:cevns_bsm} (a), $5$ MeV in Fig. \ref{fig:cevns_bsm} (b) and $10$ MeV in Fig. \ref{fig:cevns_bsm} (c). Here, we consider individual contribution of each model and also the interference effect with SM for the universal vector mediator and the $U(1)'$ models. From these figures, we may conclude as follows:
\begin{itemize}
	\item The new physics from the scalar mediator has a larger event rate than the SM in all energy scales for the chosen benchmarks. This behavior is anticipated from its cross-section that is proportional to $T_{nr}^{-1}$. This type of light mediator is then a universal way to improve measurements of low energy events. Also for heavier scalar mediators, the spectrum becomes peaked; the cross section in this case scales as $T_{nr}$ for small recoil energies, before cutting off at high energy due to a loss of coherency.
	
	\item The vector mediator contribution is generally observed when $T_{nr}$ is lower than $1$ keV. Decreasing the vector mediator mass begins to distort the spectrum. Furthermore, we can see some cancellation with the SM spectrum at a certain $T_{nr}$ due to the interference term.  We note a dip in the rate for certain recoil energies, arising from a cancellation in the $Z'$ couplings. In massive mediator scales, the cancellation effect due to the interference is unobserved, yet reducing the predicted event. This behavior indicates the importance of advancing low-energy scale measurements in search of new physics novelty.
	
	\item New physics from the tensor mediator is witnessed at low $T_{nr}$ scales, namely lower than $1$ keV, for all the mediator mass scales. This behavior is similar to the vector case with a relatively lower contribution due to the difference in the kinematic factor.
		
	\item In the $U(1)'$ models,  the $Z'$ contribution have observable effects at small recoil energy region for light mass scales. It can be seen that in high nuclear recoil energy scales, their contributions are hardly separated from the SM case. In general, these models give a higher spectrum than the universal vector case. It is due to the different charges of the fermions that determine their contributions to CE$\nu$NS where the interactions are mediated by the $Z'$ vector boson.These contributions add coherently to the SM weak neutral current interaction which is mediated by the $Z$ vector boson.
\end{itemize}

From the chosen masses, we observe that lighter ones enhance the new physics interaction spectrum, while suppress the rates as they become larger.

\begin{figure*}[htb]
	\centering
	\includegraphics[scale=0.44]{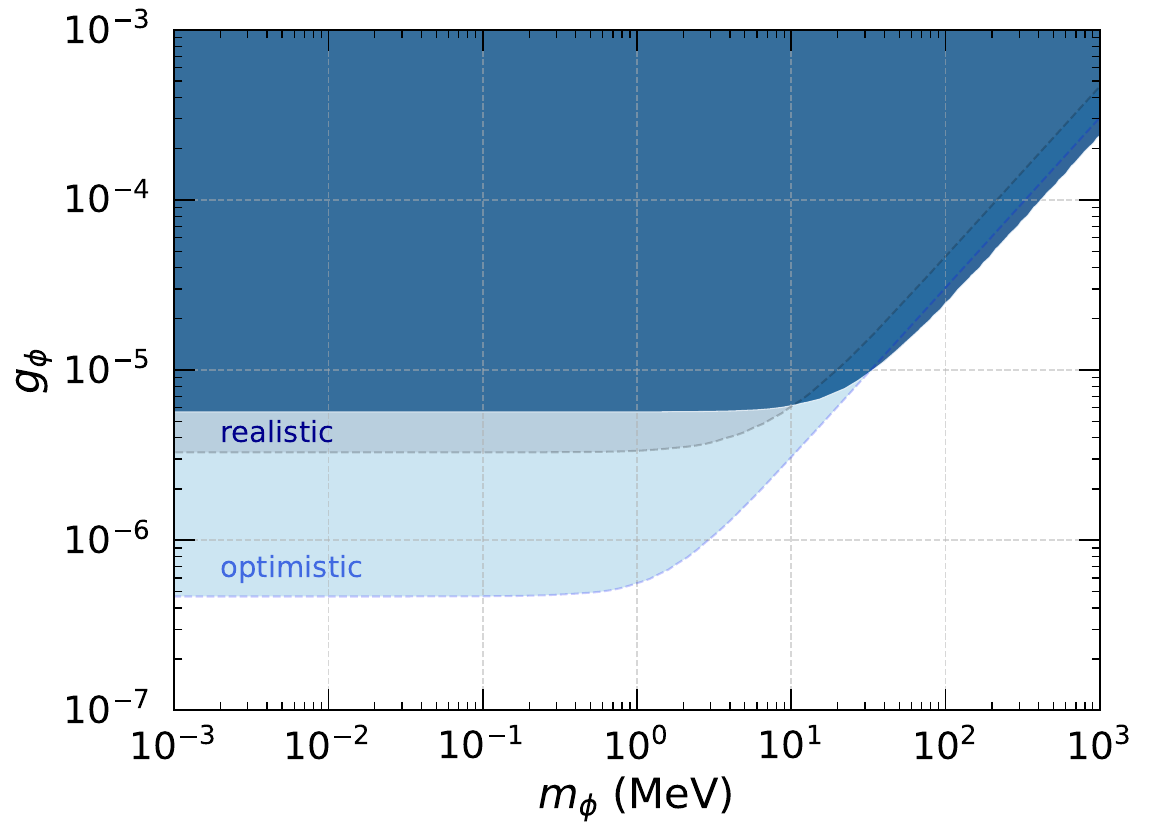}
	\includegraphics[scale=0.44]{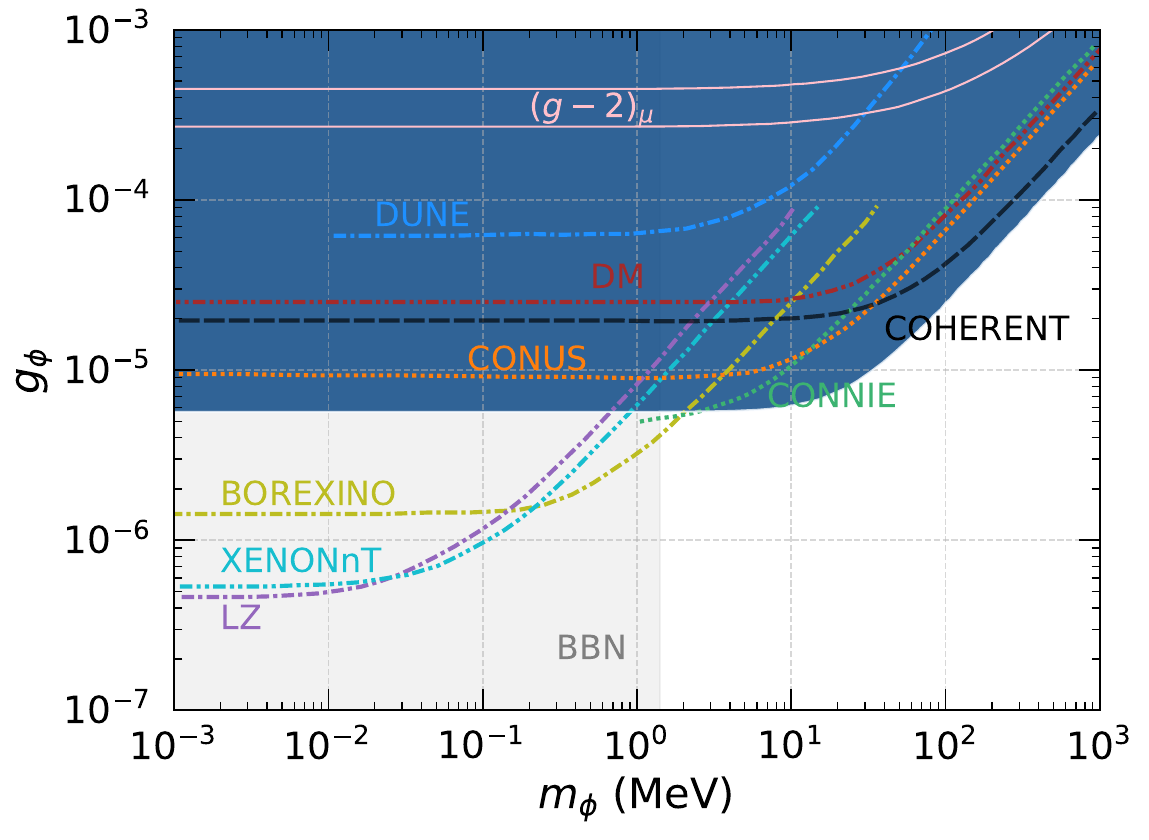}
	\\
 \hspace{10.4mm}	(a) \hspace{80.8mm} (b)
	\caption{(a) 90\% C.L. (2 d.o.f.) exclusion regions on the mass-coupling plane of the universal light scalar mediator from CE$\nu$NS by using current CDEX-10 data, and (b) comparison with other available experimental constraints. The details are in the text.
	}
	\label{fig:analysis_S}
\end{figure*}
\begin{figure*}[htb]
	\centering
	\includegraphics[scale=0.44]{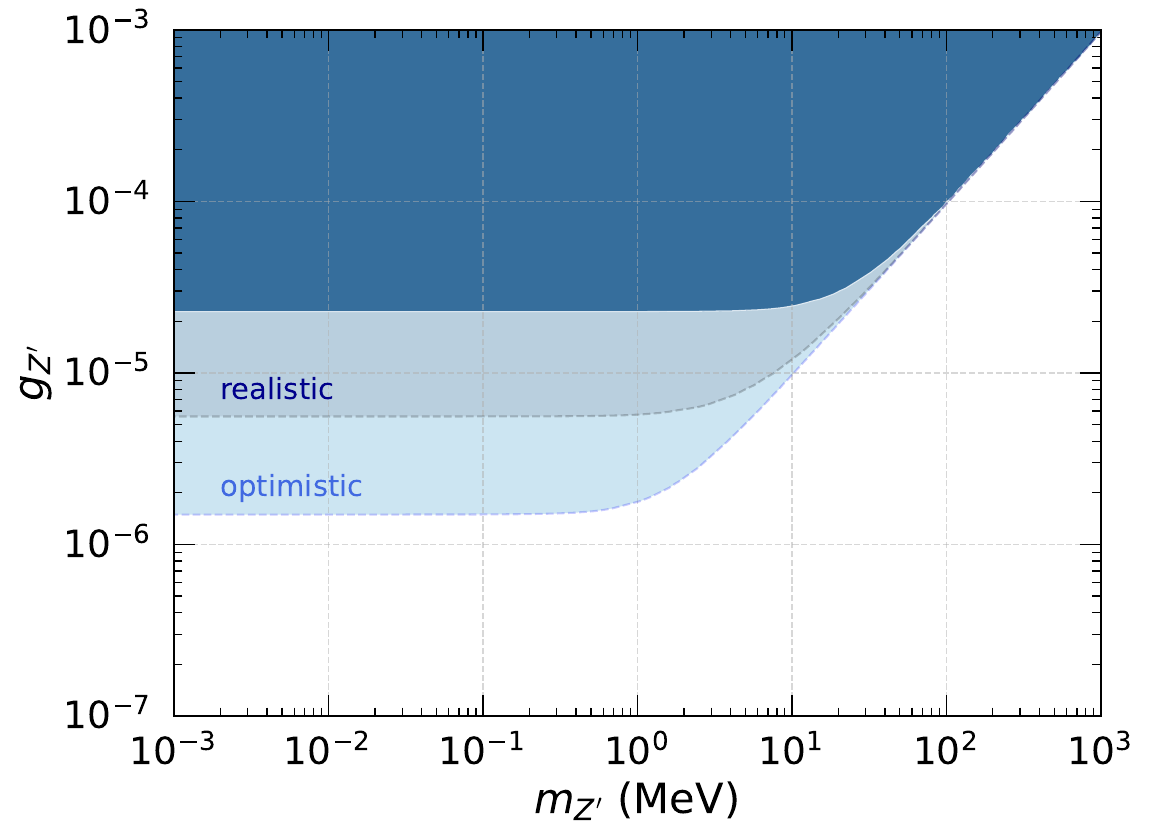}
	\includegraphics[scale=0.44]{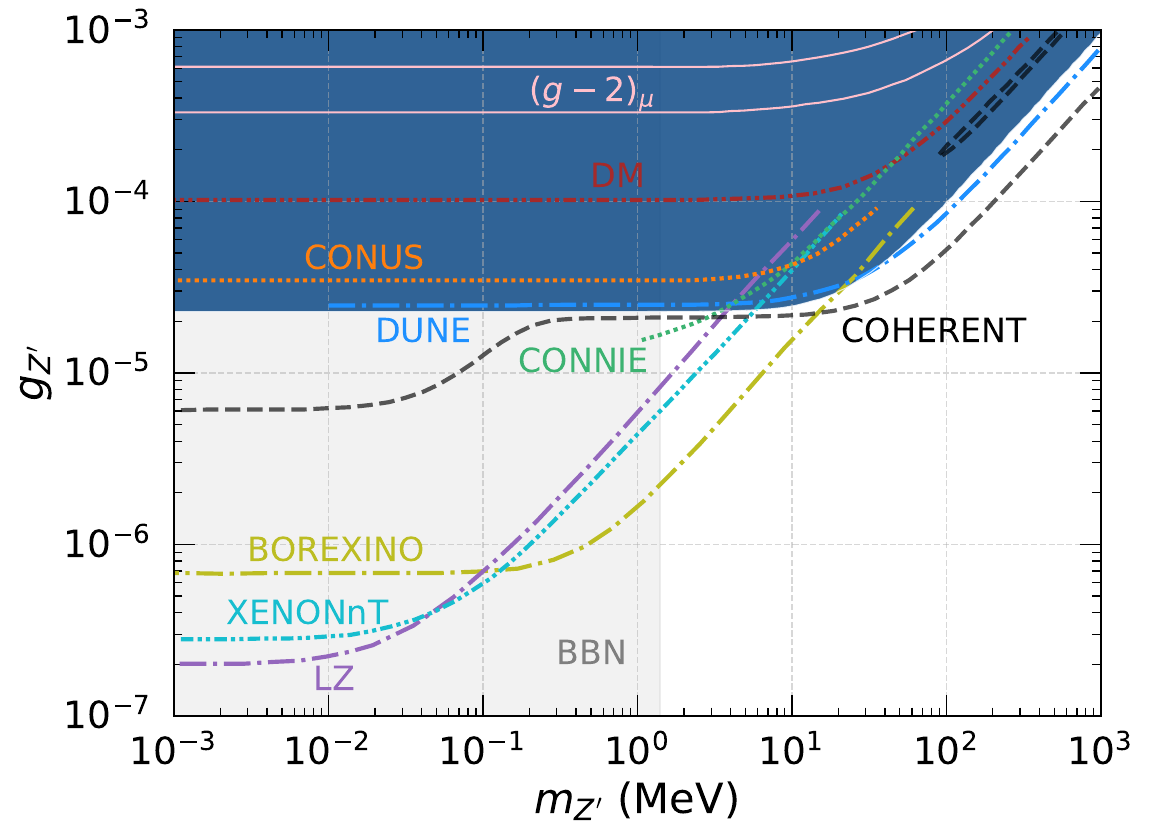}
	\\
 \hspace{10.4mm}	(a) \hspace{80.8mm} (b)
	\caption{(a) 90\% C.L. (2 d.o.f.) exclusion regions on the mass-coupling plane of the universal light vector mediator from CE$\nu$NS by using current CDEX-10 data, and (b) comparison with other available experimental constraints. The details are in the text.}
	\label{fig:analysis_V}
\end{figure*}
\subsection{Constraints on the Universal Light Mediators}
\begin{figure*}[htb]
	\centering
	\includegraphics[scale=0.44]{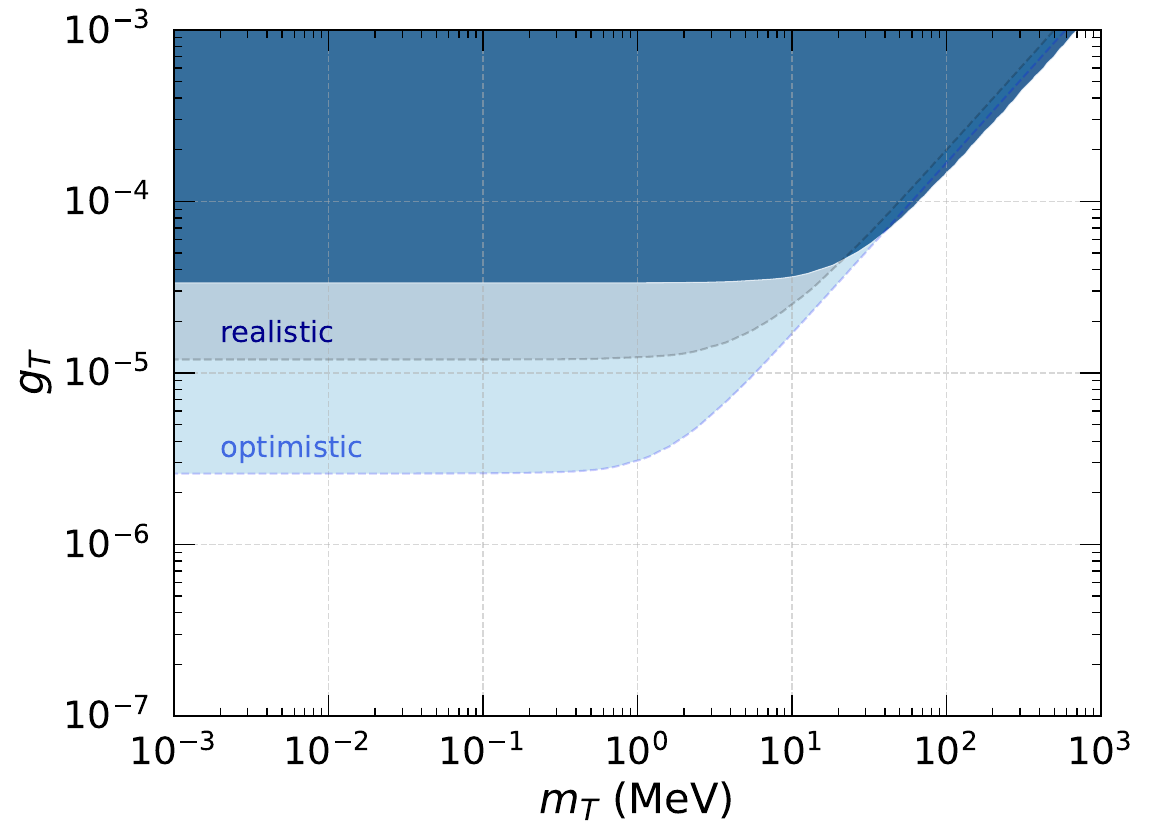}
	\includegraphics[scale=0.44]{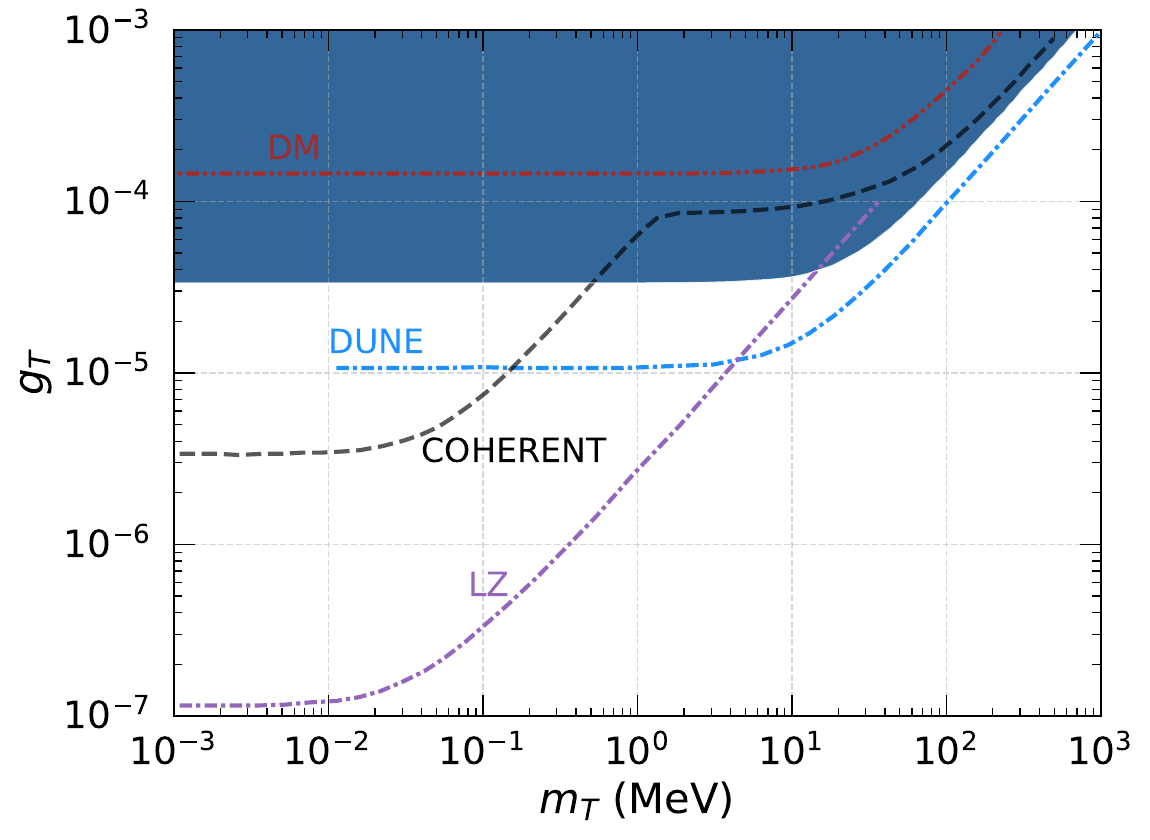}
		\\
 \hspace{10.4mm}	(a) \hspace{80.8mm} (b)
	\caption{(a) 90\% C.L. (2 d.o.f.) exclusion regions on the mass-coupling plane of the universal light tensor mediator from CE$\nu$NS by using current CDEX-10 data, and (b) comparison with other available experimental constraints (see the text) .}
	\label{fig:analysis_T}
\end{figure*}

We present the exclusion regions on the coupling-mass space of the universal light scalar, vector, and tensor mediator models in Figs. \ref{fig:analysis_S}, \ref{fig:analysis_V}, and \ref{fig:analysis_T}, respectively. We derive the constraints with $90 \%$ C.L. from 2 degrees of freedom (d.o.f.) analysis of the CDEX-10 data. We show the exclusion region from our analysis (dark blue shaded region) by superimposing our projected sensitivities from realistic and optimistic scenarios (in the left panel), and available constraints from various experimental probes (in the right panel of figures), separately.
We compare our results with current bounds derived from the different neutrino sources such as $\pi$-DAR at COHERENT with CsI+Ar target \cite{DeRomeri:2022twg} as well as nuclear reactor at CONNIE \cite{CONNIE:2019xid} and CONUS \cite{CONUS:2021dwh}. We also show the limits obtained for the CE$\nu$NS process from dark-matter experiments such as LZ \cite{LZ:2018qzl} and XENONnT \cite{A:2022acy} as well as projected dark matter \cite{Majumdar:2021vdw}. Furthermore, we include constraints obtained for the neutrino-electron scattering process from DUNE and BOREXINO  \cite{Melas:2023olz} experiments. We also show the limit from Big Bang Nucleosynthesis (BBN) \cite{Blinov:2019gcj} in the low mass regime and the $2\sigma$ allowed region of the muon anomalous magnetic moment $(g-2)_\mu$
\cite{Muong-2:2023cdq}.

For the universal scalar mediator, it can be seen from Fig. \ref{fig:analysis_S}(a) that the upper-limit of the coupling constant $g_\phi$ is about $5.68 \times 10^{-6}$ in the region of $m_\phi < 0.8 $ MeV. We also show the projected sensitivities as indicated from the dark blue to the lighter blue region. The realistic and optimistic scenarios yield an improvement of approximately $42.3\%$ and $91.8\%$ over the current constraint, corresponding to $g_\phi\lesssim 3.28\times 10^{-6}$ and $g_\phi\lesssim 4.67\times 10^{-7}$, respectively.

%
%
\begin{table*}[ht]
	\caption{90\% C.L. (2 d.o.f.) upper-limits on the coupling constants for the universal light mediator models.
Here, we give the lowest-limit values from COHERENT and XENONnT in the considered parameter region.
	}
	\begin{center}
		\begin{ruledtabular}
		\begin{tabular}{ c c c c c c  }
			\multirow{2}{1.5cm}{Coupling} & \multicolumn{3}{c}{CDEX (\textbf{this work})} & \multirow{2}{2cm}{COHERENT} &  \multirow{2}{2cm}{XENONnT} \\
			\cline{2-4}
			& Current & Realistic & Optimistic  &   &  
			\\
			\hline
			$g_\phi$ & $\lesssim 5.68\times 10^{-6}$ & $ \lesssim 3.28\times 10^{-6}$ & $ \lesssim 4.67\times 10^{-7}$ &
			$ \lesssim 1.96\times 10^{-5}$ & $ \lesssim 5.34\times 10^{-7}$
			\\

			 $g_{Z'}$ & $\lesssim 2.28\times 10^{-5}$ &  $ \lesssim 5.57\times 10^{-6}$ & $ \lesssim 1.49\times 10^{-6}$ &
			$ \lesssim 6.08\times 10^{-6}$ &
			$ \lesssim 2.81\times 10^{-7}$  
			\\
			$g_{T}$ & $ \lesssim 3.35\times 10^{-5}$ & $ \lesssim 1.20\times 10^{-5}$ & $ \lesssim 2.59\times 10^{-6}$ &
			$\lesssim 3.37\times 10^{-6}$ &
			- 
		    \\
		\end{tabular}
			\end{ruledtabular}
	\end{center}
	\label{tab:couplings_univ}
\end{table*}

We overlaid limits of previous studies outlined above in Fig. \ref{fig:analysis_S}(b). It is clear that our result provides a more stringent constraint than those obtained from COHERENT, CONNIE, CONUS, DUNE and projected DM.  However, the results from BOREXINO, LZ, and XENONnT experiments are not fully covered at low mass regions of $m_\phi < 2.0$ MeV, $m_\phi < 0.7$ MeV, and $m_\phi < 0.9$ MeV, respectively. As for the BBN, the current study reaches the limit as $m_\phi \leq 1.4$ MeV. We also indicate the preferred region of $(g-2)_\mu$ with $2\sigma$ that is fully covered. Considering our projected scenarios, all limits from previous works for the scalar mediator model would be reachable in future developments.

For the universal light vector mediator, our constraint together with projected sensitivities are shown in Fig. \ref{fig:analysis_V}(a). The upper-limit of $g_{Z'}$ reaches a value of about $2.28 \times 10^{-5}$ in the region of $m_{Z'}<1.3$ MeV. The realistic and optimistic scenarios yield an improvement of approximately $75.6\%$ and $93.5\%$ over the current constraint, which correspond to  $g_{Z'} \lesssim 5.57\times 10^{-6}$ and $g_{Z'} \lesssim 1.49\times 10^{-6}$, respectively.

In Fig. \ref{fig:analysis_V}(b), we superimpose for the universal vector mediator other avaliable constraints outlined above. It is seen that our result provides a more stringent constraint than those obtained from CONUS and projected DM. Limit of CONNIE is reached down to around $m_{Z'}<3.4$ MeV while the bound of DUNE is covered only in the low mass region of $m_{Z'} < 35$ MeV.
The LZ, XENONnT and BOREXINO limits are reached as the mediator mass is  $m_{Z'}> 3.8 $ MeV, $m_{Z'} > 5.8$ MeV, and $m_{Z'} > 21 $ MeV, respectively. 
The region of $(g-2)_\mu$ is entirely covered. 
Furthermore, the combination of CsI and Ar from the COHERENT limit is yet to be reached.
It is anticipated that the projected scenarios could reach the low mass limit of the COHERENT, but yet to cover all exclusion regions of BOREXINO, LZ, and XENONnT.

In Fig. \ref{fig:analysis_T}(a), we show our constraint together with projected sensitivities for the universal light tensor mediator. The upper-limit of $g_T$ reaches a value of about $3.35\times 10^{-5}$ in the region of $m_T < 0.8$ MeV. The realistic and optimistic scenarios yield an improvement of around $64.2\%$ and $92.3\%$ over the current constraint, which correspond to  $g_T\lesssim 1.20\times 10^{-5}$ and $g_T\lesssim 2.59\times 10^{-6}$, respectively.

We overlaid the existing limits from the previous studies of the universal light tensor mediator in Fig. \ref{fig:analysis_T}(b). It is seen that the limit of the projected DM is all covered. Moreover, our result shows a more competitive constraint than those obtained from COHERENT in the mass region of $m_T > 0.5$ MeV. The LZ limit is reached as the mediator mass is $m_T > 14$ MeV. Meanwhile, the DUNE constraint is yet to be covered. For the projected scenarios, they are able to fully cover the limit of COHERENT and partially reach limits of DUNE and LZ in the light mass scale.

\begin{figure*}[htb]
	\centering
	\includegraphics[scale=0.44]{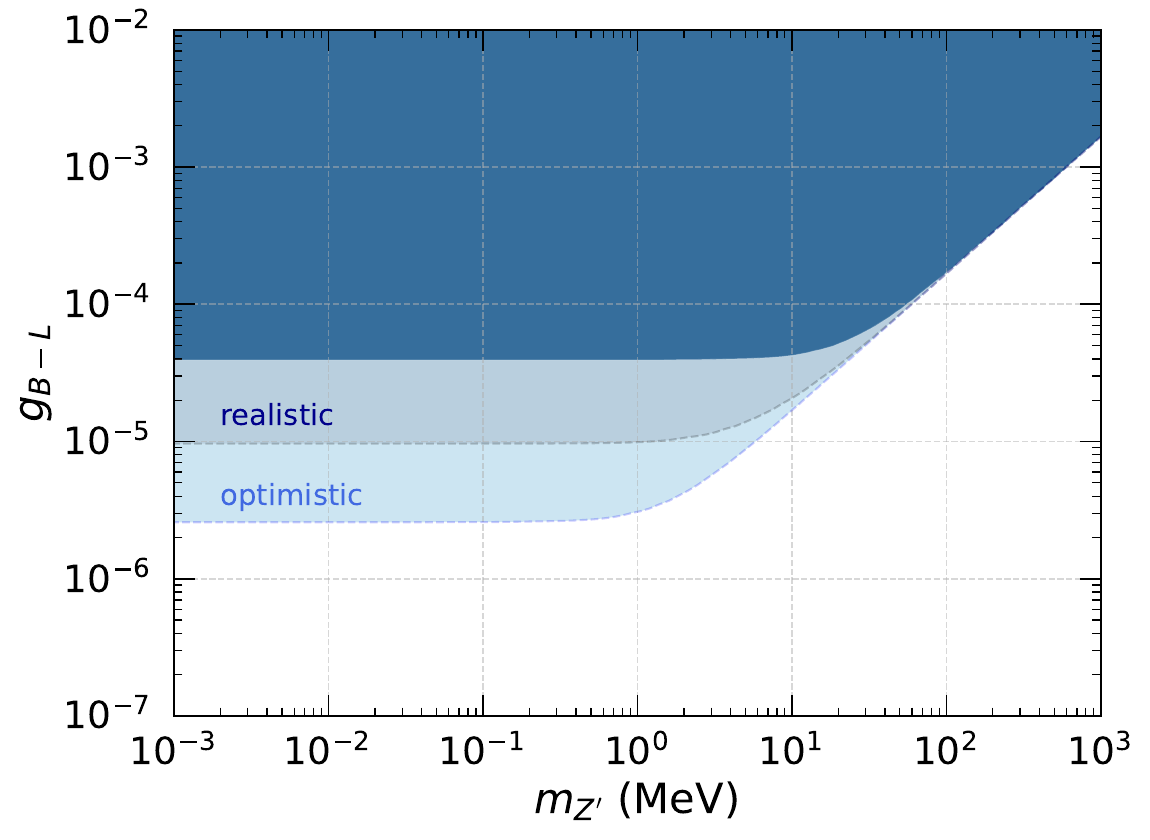}
	\includegraphics[scale=0.44]{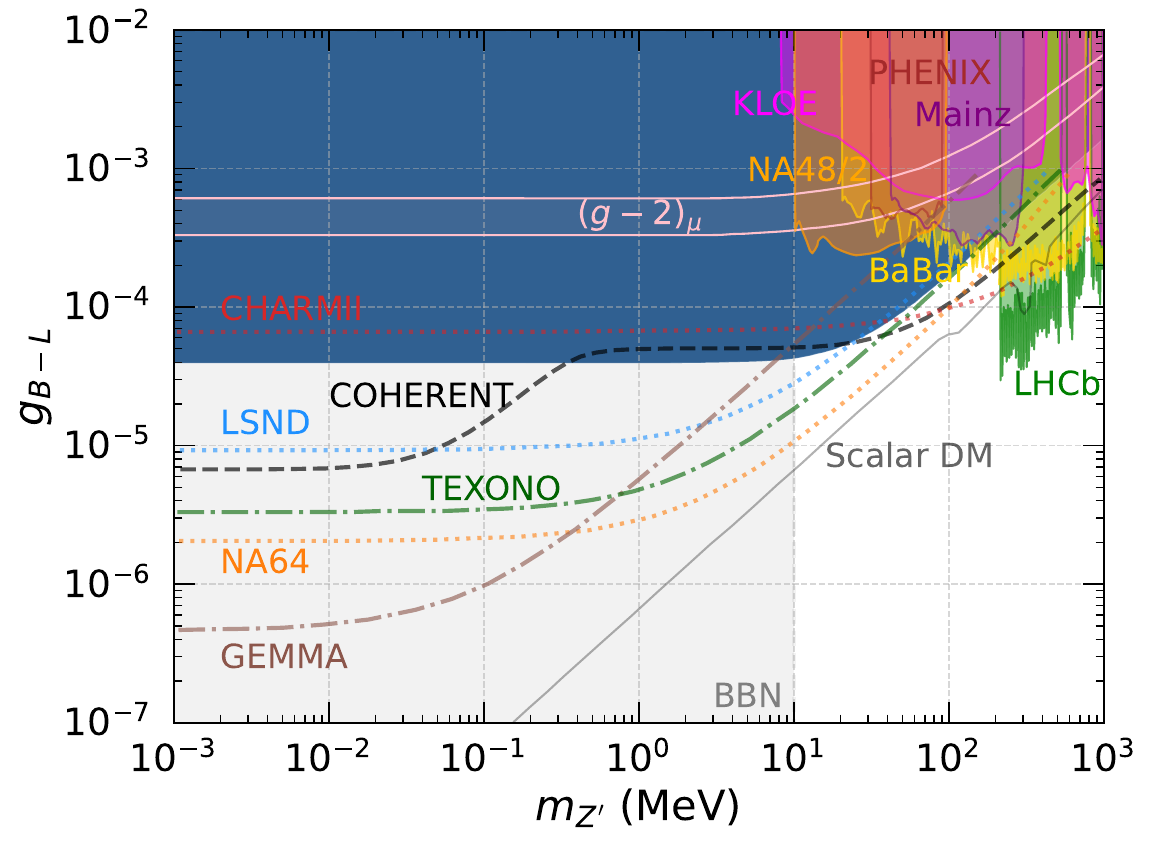}
	\\
 \hspace{10.4mm}	(a) \hspace{80.8mm} (b)
	\caption{(a) 90\% C.L. (2 d.o.f.) exclusion regions on the mass-coupling plane of the $U(1)_{B-L}$ model from CE$\nu$NS by using current CDEX-10 data, and (b) comparison with other available experimental constraints (see the text for details).}
	\label{fig:analysis_BL}
\end{figure*}
Our results generally yield an updated constraint according to existing limits from previous studies. We summarize our results in Table \ref{tab:couplings_univ}. One can read these limits directly from Figs.  \ref{fig:analysis_S}, \ref{fig:analysis_V} and \ref{fig:analysis_T}. The COHERENT bound is outperformed in the scalar case, partly covered in the tensor case, and yet to be reached in the vector case. Constraints from reactor studies, namely CONUS and CONNIE, as well as the $(g-2)_\mu$ are all covered. The limits from CDEX-10 are competitive with the projected DM in all of the mediator types. Meanwhile, the LZ, BOREXINO, and XENONnT limits dominate at low mass regions since they are capable of detecting very low-energy thresholds.  Finally, the current work covers the exclusion region of DUNE in the universal scalar mediator model, slightly dominant in the vector case, and outperformed in the tensor case.

\subsection{Constaints on the $U(1)'$ models}
\subsubsection{$B-L$}
In Fig. \ref{fig:analysis_BL}(a), we show the $90\%$ C.L excluded region on the coupling-mass plane of the $B-L$ model, including projected sensitivities from realistic and optimistic scenarios.
As expected, the constraints of ${B-L}$ model cover a smaller region than the universal vector mediator model, due to the $U(1)'$ charge factor. Our analysis indicates that CDEX-10 data improve previous results in some region.  The upper-limit of $g_{B-L}$ reaches a value of $3.93 \times 10^{-5}$ in the region of $m_{Z'}< 0.1$ MeV. The realistic and optimistic scenarios yield improvements of approximately $75.4\%$ and $93.4\%$ over the current constraint, which correspond to  $g_{B-L} \lesssim 9.66\times 10^{-6}$ and $g_{B-L} \lesssim 2.58\times 10^{-6}$, respectively.

\begin{figure*}[htb]
	\centering
	\includegraphics[scale=0.44]{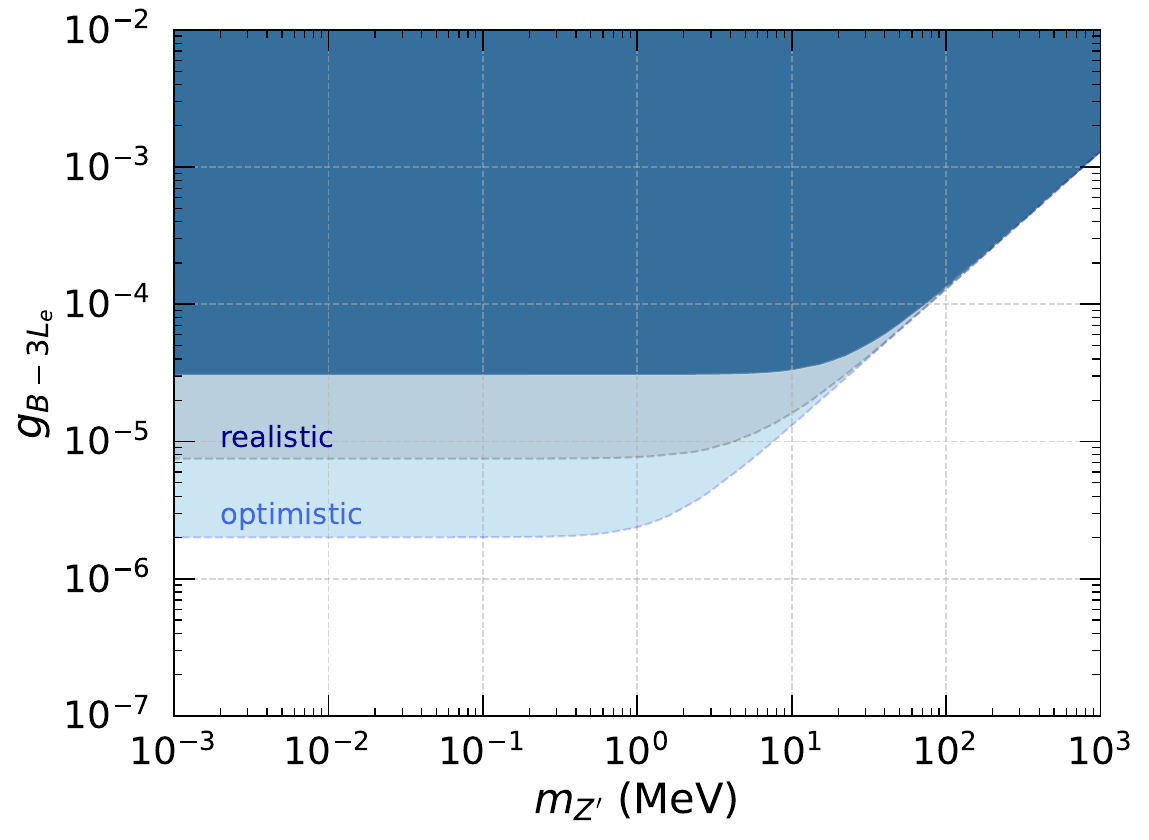}
	\includegraphics[scale=0.44]{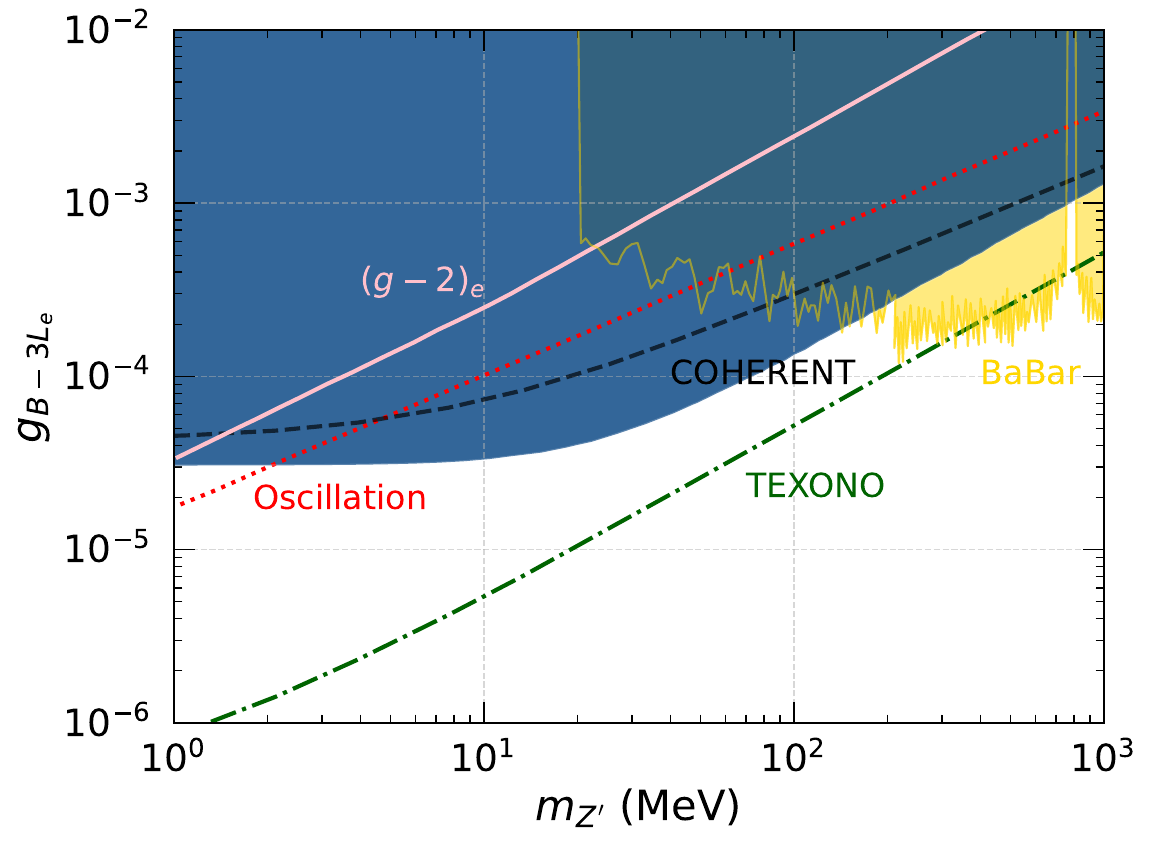}
	\\
	\hspace{10.4mm}	(a) \hspace{80.8mm} (b)
	\caption{(a) 90\% C.L. (2 d.o.f.) exclusion regions on the mass-coupling plane of the $U(1)_{B-3L_e}$ model from CE$\nu$NS by using current CDEX-10 data, and (b) comparison with other available experimental constraints (see the text for details).}
	\label{fig:analysis_B3Le}
\end{figure*}
\begin{figure*}[htb]
	\centering
	\includegraphics[scale=0.44]{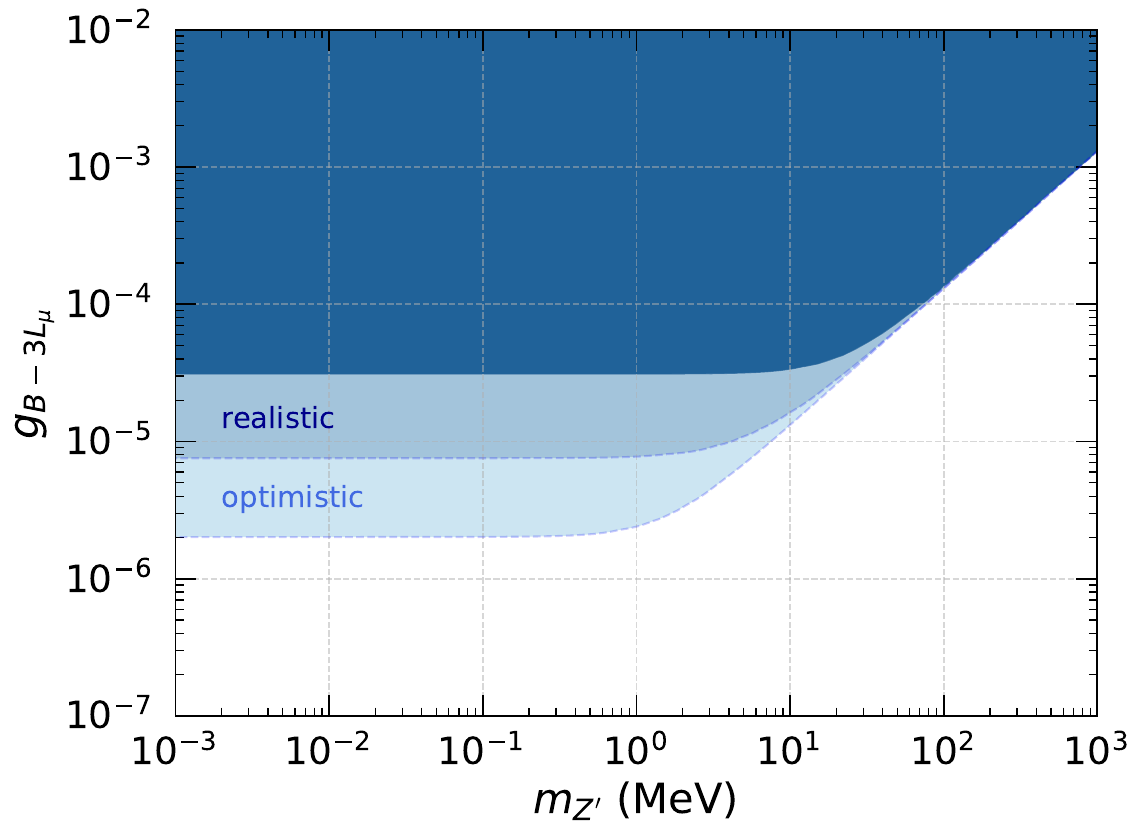}
	\includegraphics[scale=0.44]{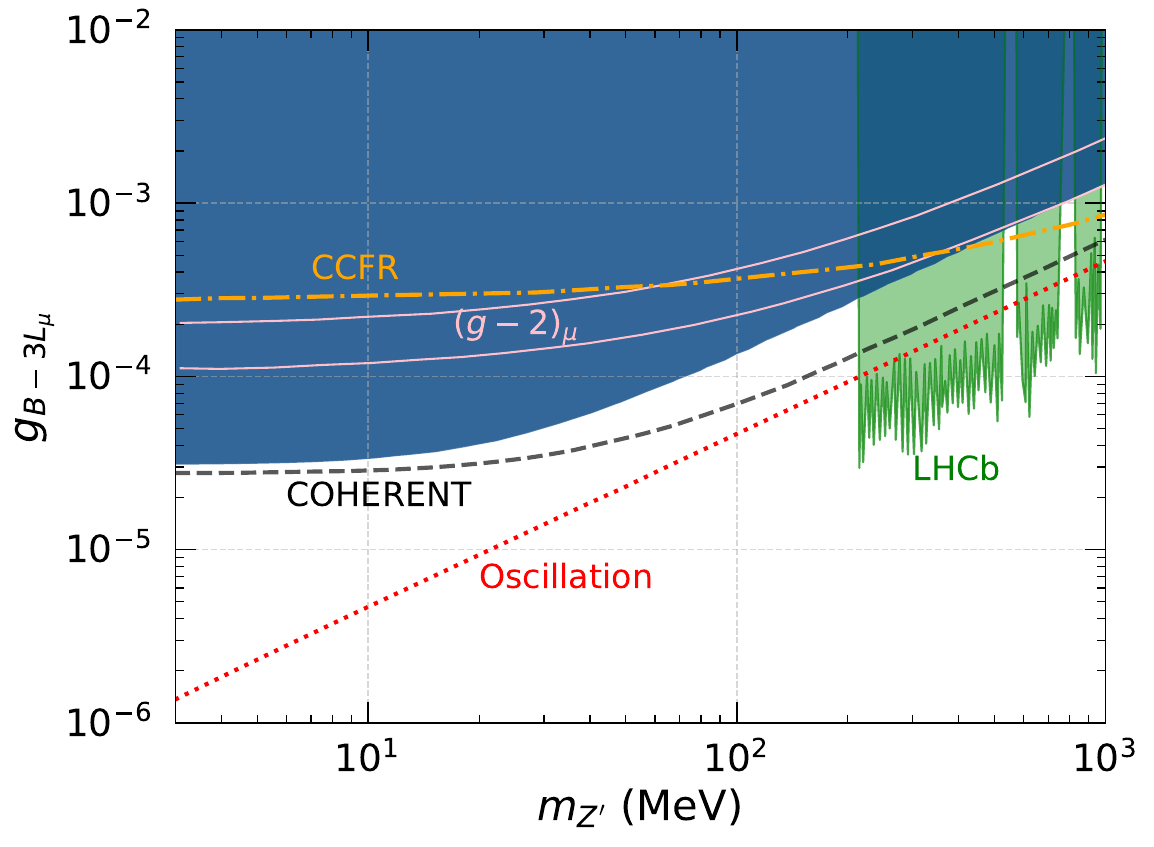}
	\\
	\hspace{10.4mm}	(a) \hspace{80.8mm} (b)
	\caption{(a) 90\% C.L. (2 d.o.f.) exclusion regions on the mass-coupling plane of the  $U(1)_{B-3L_\mu}$ model from CE$\nu$NS by using current CDEX-10 data, and (b) comparison with other available experimental constraints (see the text for details).}
	\label{fig:analysis_B3Lm}
\end{figure*}
In Fig. \ref{fig:analysis_BL}(b), we overlaid our analysis of the $B-L$ model with available limits obtained from CE$\nu$NS experiment (COHERENT with CsI+Ar target, derived in Ref. \cite{DeRomeri:2022twg}), neutrino-electron scattering experiments (LSND \cite{LSND:2001akn}, CHARMII \cite{CHARM-II:1994dzw}, and NA64 \cite{NA64:2022yly}), nuclear reactor experiments (TEXONO \cite{TEXONO:2009knm} and GEMMA \cite{Beda:2009kx}), collider experiments (BaBar \cite{BaBar:2014zli}, LHCb\cite{LHCb:2019vmc}, KLOE \cite{ALICE:2012aqc}, Mainz \cite{A1:2011yso}, PHENIX \cite{PHENIX:2014duq}), and rare-meson decay experiment (NA48/2 \cite{NA482:2015wmo}). We also show the $(g-2)_\mu$ bound \cite{Muong-2:2023cdq} as well as scalar thermal DM \cite{Berlin:2018bsc}. The current data provide more stringent results than the COHERENT limit in the region of $0.34 \text{ MeV}\leq m_{Z'} \leq 23 \text{ MeV}$.
The CHARMII bound is covered as the mediator mass $m_{Z'}\leq 42 \text{ MeV}$, while the GEMMA limit is covered as $m_{Z'} \geq 7.6 \text{ MeV}$, and slightly better than the LSND result for $m_{Z'} \geq 25.6 \text{ MeV}$. 
The TEXONO bound is reached as $m_{Z'} \geq 166 \text{ MeV}$, while the NA64 bound is yet to be reached.  
Moreover, the $Z'$ could mediate new feeble interaction between the scalar DM and SM particles \cite{Berlin:2018bsc}. This bound can be obtained by assuming the mass ratio of the scalar DM $\chi$ with the $Z'$ of the $B-L$ model to be $m_{Z'}=3m_\chi$ \cite{NA64:2022yly}. This bound is yet to be reached by the current data. Furthermore, the BBN limit is reached as $m_{Z'}<10 \text{ MeV}$.
Regarding the constraints from colliders, our analysis indicates that CDEX-10 data can cover parameter-space for a mass range between $8.2$ MeV to $1.0$ GeV.

\subsubsection{$B-3L_e, B-3L_\mu$, and $B-3L_\tau$}
\begin{figure*}[htb]
	\centering
	\includegraphics[scale=0.44]{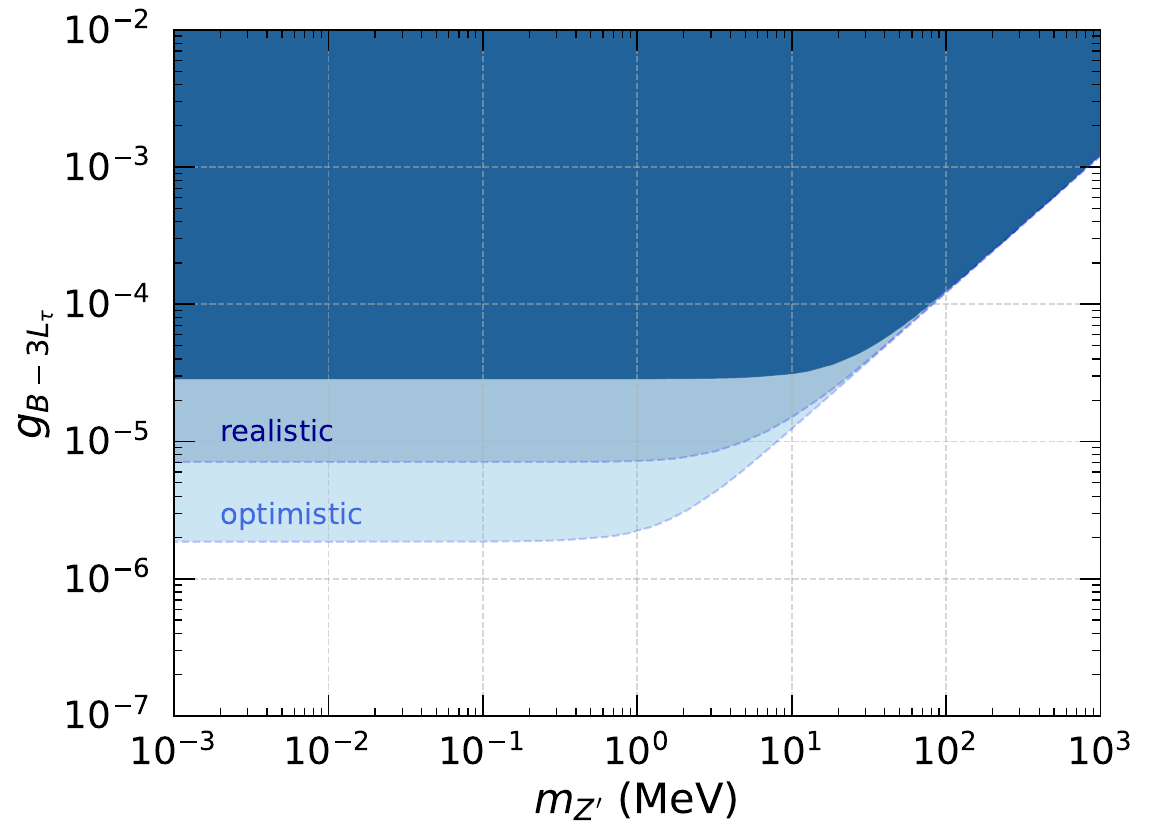}
	\includegraphics[scale=0.44]{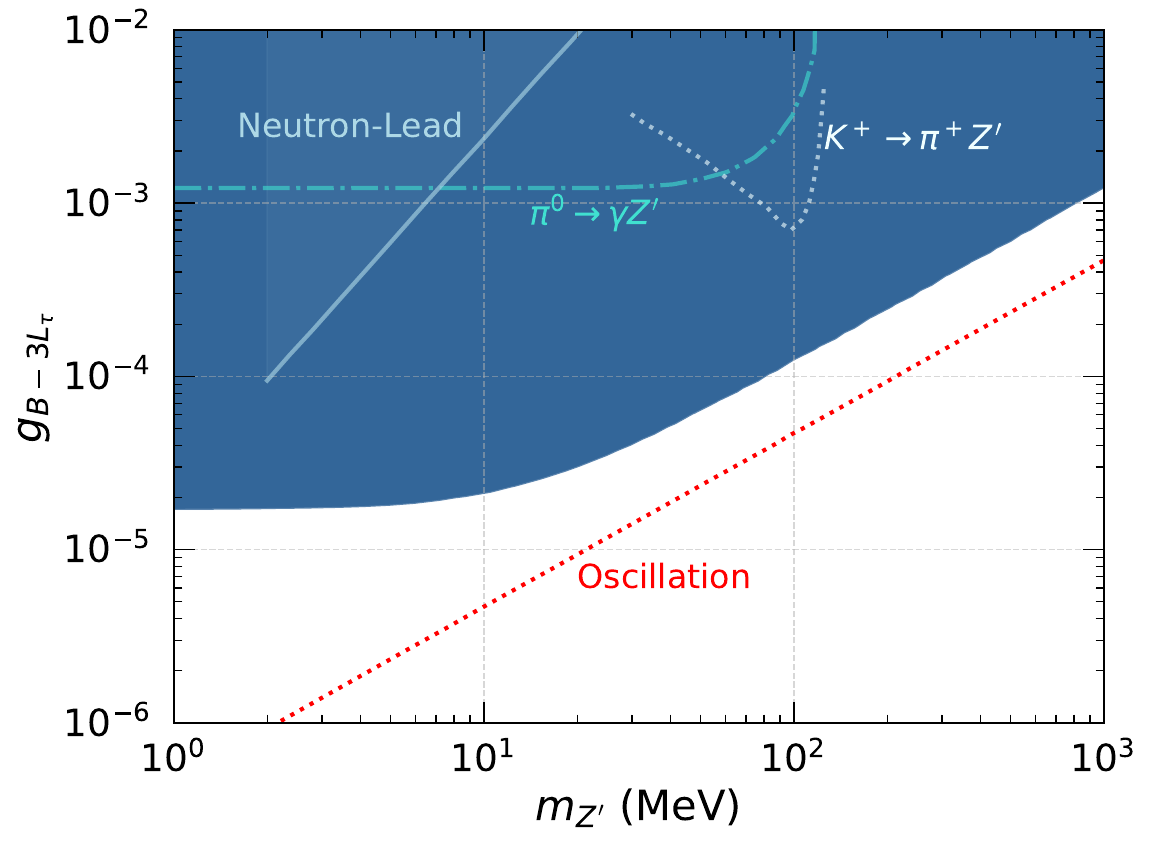}
	\\
 \hspace{10.4mm}	(a) \hspace{80.8mm} (b)
	\caption{(a) 90\% C.L. (2 d.o.f.) exclusion regions on the mass-coupling plane of the  $U(1)_{B-3L_\tau}$ model from CE$\nu$NS by using current CDEX-10 data, and (b) comparison with other available experimental constraints (see the text for details).}
	\label{fig:analysis_B3Lt}
\end{figure*}

In Figs. \ref{fig:analysis_B3Le}, \ref{fig:analysis_B3Lm}, and \ref{fig:analysis_B3Lt}, we show the $90\%$ C.L. excluded region from CDEX-10 data on the coupling-mass plane of the $B-3L_e$, $B-3L_\mu$ and $B-3L_\tau$ models, including projected sensitivities from realistic and optimistic scenarios.
We present the current limits obtained from previous studies for comparison. These are derived from several neutrino experiments such as $\pi$-DAR (COHERENT \cite{AtzoriCorona:2022moj}), nuclear reactor (TEXONO \cite{Heeck:2018nzc}), colliders (BaBar \cite{BaBar:2014zli} and LHCb \cite{LHCb:2019vmc}), neutrino tridents (CCFR \cite{CCFR:1991lpl}), as well as neutron-lead scattering \cite{Barbieri:1975xy}. We also include the limits derived from the global analysis of oscillation data for each model \cite{Coloma:2020gfv}. In addition, results of $(g-2)_\mu$ and $(g-2)_e$ \cite{AtzoriCorona:2022moj} are also included. For the $B-3L_\tau$ model, we include bound prediction from pion and kaon decays \cite{Heeck:2018nzc}. In general, our results provide a more stringent limit than most of the mentioned works while yet to reach others. We give our evaluations separately for each model in the following.

In Fig. \ref{fig:analysis_B3Le}(a), we present the excluded region and projected sensitivities for the $B-3L_e$ model. The current data yields an upper-limit of $g_{B-3L_e}$ as $3.06\times10^{-5} $ for the mass region of $m_{Z'}< 1.6$ MeV. The realistic and optimistic scenarios yield an improvement of approximately $75.4\%$ and $93.4\%$ over the current constraint, which correspond to $g_{B-3L_e}\lesssim 7.50\times 10^{-6} $ and $g_{B-3L_e}\lesssim 2.00 \times 10^{-6} $, respectively. Comparison with other limits is shown in Fig. \ref{fig:analysis_B3Le}(b). Here, the COHERENT and  $(g-2)_e$  limits are fully covered, while the oscillation  limit is reached as $m_{Z'}< 2.07\text{ MeV}$ within the considered parameter space. Furthermore, it also mainly covers the limit of BaBar, while the bound of TEXONO is yet to be reached.

\begin{table*}[ht!]
	\caption{The same as Table \ref{tab:couplings_univ} but for $U(1)'$ models.  
	We also present the lowest-limit of COHERENT in the considered parameter region.}
	\begin{center}
	\begin{ruledtabular}
		\begin{tabular}{ l c c c c}
			\multirow{2}{1.5cm}{Coupling} & \multicolumn{3}{c}{CDEX (\textbf{this work})} & \multirow{2}{2cm}{COHERENT} 
			\\
			\cline{2-4}
			& Current & Realistic & Optimistic  &   
			\\
			\hline
			$g_{B-L}$ & $\lesssim 3.93\times 10^{-5}$ & $ \lesssim 9.66\times 10^{-6}$ & $ \lesssim 2.58\times 10^{-6}$ & $\lesssim 6.74\times 10^{-6}$ 
			\\
			$g_{B-3L_e}$ & $\lesssim 3.06\times 10^{-5}$ & $ \lesssim 7.50\times 10^{-6}$ & $ \lesssim 2.00\times 10^{-6}$ &
			$\lesssim 4.54\times 10^{-5}$ 
			\\
			$g_{B-3L_\mu}$ & $\lesssim 3.07\times 10^{-5}$ & $ \lesssim 7.55\times 10^{-6}$ & $ \lesssim 2.01\times 10^{-6}$ &
			$\lesssim 2.76\times 10^{-5}$ 
			\\
			$g_{B-3L_\tau}$ & $\lesssim 2.84\times 10^{-5}$ & $ \lesssim 7.07\times 10^{-6}$ & $ \lesssim 1.86\times 10^{-6}$ &
			- 
			\\
		\end{tabular}
		\end{ruledtabular}
	\end{center}
	\label{tab:couplings_u1}
\end{table*}

In Fig. \ref{fig:analysis_B3Lm}(a), we show the excluded region and projected sensitivities for $B-3L_\mu$ model. The current data provides an upper-limit of $g_{B-3L_\mu}$ as $3.07\times10^{-5} $ for the mass region of $m_{Z'}< 0.1$ MeV. The realistic and optimistic scenarios yield an improvement of approximately $75.4\%$ and $93.4\%$ over the current constraint, which correspond to $g_{B-3L_\mu}\lesssim 7.55\times 10^{-6}$ and $g_{B-3L_\mu}\lesssim 2.01\times 10^{-6}$, respectively. Figure \ref{fig:analysis_B3Lm}(b) shows a comparison with other limits. Bound results of $(g-2)_\mu$ and CCFR are mainly covered, while the ones from the COHERENT as well as oscillation are yet to be reached, and the LHCb constrain is partially covered with the current analysis.

Finally, we present the excluded region and projected sensitivities for the $B-3L_\tau$ model in Fig. \ref{fig:analysis_B3Lt}(a). The current data yields an upper-limit of $g_{B-3L_\tau}$ as $2.84\times10^{-5} $ for the mass region of $m_{Z'}< 0.9$ MeV. The realistic and optimistic scenarios yield an improvement of approximately $75.1\%$ and $93.4\%$ over the current constraint, which correspond to $g_{B-3L_\tau}\lesssim 7.07\times 10^{-6}$ and $g_{B-3L_\tau}\lesssim 1.86\times 10^{-6}$, respectively.
In Fig. \ref{fig:analysis_B3Lt}(b), we overlaid the available limits from previous works for comparison.
It is obvious that the current study could explain the limits from the kaon and pion decays, as well as neutron-lead scattering. Meanwhile, the oscillation limit is yet to be reached.

Overall, our results indicate competitive constraints compared to some existing limits from previous studies. We summarize the upper-limit findings in Table \ref{tab:couplings_u1}. One can read these limits directly from Figs.  \ref{fig:analysis_BL}, \ref{fig:analysis_B3Le}, \ref{fig:analysis_B3Lm} and \ref{fig:analysis_B3Lt}. The COHERENT bounds are outperformed in the $B-3L_e$ model for the considered mass region while in the $B-3L_\mu$ model it still yet to be reached by this current analysis. The $(g-2)_e$ limit is all covered, whereas the $(g-2)_\mu$ limit is slightly more stringent in high mass regions. The TEXONO result outperforms the CDEX-10 limit for the $B-3L_e$ model, while the CCFR limit is competitive in the massive mediator region for the $B-3L_\mu$ model. As for collider results of BaBar and LHCb, both are partially covered and competitive in high mass scale. Furthermore, it is seen from the analysis of the $B-3L_\tau$ model that the CDEX-10 limit dominates those from neutron-lead scattering, as well as the pion and kaon decays. Finally, the oscillation limits are mainly covered in the $B-3L_e$ model, while in both the $B-3L_\mu$ and $B-3L_\tau$ models they outperform the current analysis.

\section{Summary and Conclusions}\label{sec:summ}
We have studied new physics from the light mediator models of the universal light mediators and the $U(1)'$ symmetries through the process of CE$\nu$NS with solar neutrino using recent CDEX-10 data. Accordingly, we have focused on general scalar, vector, and tensor interactions allowed by Lorentz invariance and involving universal light mediators. 
Furthermore, we have considered an additional vector gauge boson with associated $U(1)'$ gauge group for variety of models including $U(1)_{B-L}$, $U(1)_{B-3L_e}$, $U(1)_{B-3L_\mu}$, and $U(1)_{B-3L_\tau}$. These considerations could account for possible new physics effects as neutrinos interact with matter, namely quark components of nuclei.

We have calculated the CE$\nu$NS differential rate for each model using solar neutrino fluxes by concerning the quenching factor to convert between nuclear recoil energy and its electron equivalent. For this conversion, we use two different quenching factors:  Linhard quenching factor and  the ”high” ionization-efficiency model quenching factor.
 In general, these models give a higher spectrum than the SM case. The $U(1)'$ models  differ in the charges of the fermions that determine their contributions to CE$\nu$NS of the interactions mediated by the $Z'$ vector boson. These contributions add coherently to the SM weak neutral current interaction. The effects of new contributions from these models have emerged in low-scale nuclear recoil energy, indicating the need to increase detector sensitivity in this domain.

We have derived new constraints on the coupling constant-mass plane of the considered models by using recent CDEX-10 data. Furthermore, we have suggested two projection scenarios, namely realistic and optimistic scenarios, regarding the experiment's advancement, and discussed their sensitivities. We compared our results with the existing limits derived from further experimental probes. In both universal light mediators and $U(1)'$ models, our results indicate that some existing limits are mainly covered while yet to reach others.

For the universal scalar and vector mediator models, our results improve the limits derived from CONUS, CONNIE, and projected DM. For the universal tensor mediator model, the projected DM limit is also outperformed. As for the DUNE limit, our result improves all of its regions for the universal scalar mediator model, while it outperforms our result in high mass scale for the vector case, and it completely dominates for the tensor case. Concerning the COHERENT limits, our result dominates in the scalar case, yet to reach in the vector case, while it indicates partial improvements in tensor case. Limits from BOREXINO, XENONnT, and LZ (at least for a significant part of them) are yet to be reached in all the considered models. However, the projected sensitivities could dominate these limits in the universal scalar mediator model, while still being outperformed in the vector and tensor cases.

The exclusion regions on the $B-L$ model show that the current result could lead an improvement in the existing limits.
Some improvements are found for the CHARM limit in the low mass scales and for the GEMMA and the LSND limits in the high mass scales. It also improves over the limit from COHERENT in the intermediate mass scale, while still being outperformed in the low and high mass scales. Moreover, it is competitive with the TEXONO limit in the high mass region and has yet to reach the limit of NA64. Moreover, bounds from the collider studies are partially covered.
For the $B-3L_e$ model, our result dominates bounds from $(g-2)_e$, oscillation, and COHERENT, while it partially covers the BaBar limit and outperformed by the TEXONO limit. For the $B-3L_\mu$ model, there is an improvement at low mass region according to the CCFR limit. It partially covers the LHCb limit and has yet to reach the COHERENT and oscillation limits. Finally, for the $B-3L_\tau$ model, it still outperformed by oscillation limit, while it dominates limits from the neutron-lead as well as the predicted pion and kaon decays. We note that the projected sensitivities could generally dominate over these limits on $U(1)'$ models.

In conclusion, CE$\nu$NS with a solar neutrino source can provide improvements to the existing limits on light mediator models. This hence indicates the importance of combining both solar neutrino and direct detection experiments to provide a better understanding of neutrino interactions. We anticipate that our results could be used as ancillary considerations for hunting new physics with current and future experimental advancements related to solar and other neutrino sources.

\section{Acknowledgments}
We would like to thank L. T. Yang from CDEX collaboration for sharing information on the CDEX-10 data and useful correspondence.
This work was supported by the Scientific and Technological Research Council of Turkey (TUBITAK) under the project no: 123F186.


\begin{thebibliography}{99}
\bibitem{Freedman:1973yd}
D.~Z.~Freedman, Coherent Neutrino Nucleus Scattering as a Probe of the Weak Neutral Current,
\href{https://doi.org/10.1103/PhysRevD.9.1389}{{Phys. Rev. D} \textbf{9}, 1389-1392 (1974)}.


\bibitem{Akimov:2017ade}
D.~Akimov \textit{et al.} [COHERENT],
Observation of Coherent Elastic Neutrino-Nucleus Scattering,
\href{https://doi.org/10.1126/science.aao0990}{{Science} \textbf{357}, 1123-1126 (2017)}.

\bibitem{Akimov:2020pdx}
D.~Akimov \textit{et al.} [COHERENT],
First Measurement of Coherent Elastic Neutrino-Nucleus Scattering on Argon,
\href{https://doi.org/10.1103/PhysRevLett.126.012002}{{Phys. Rev. Lett.} \textbf{126} 012002, (2021)}.



\bibitem{COHERENT:2021xmm}
D.~Akimov \textit{et al.} [COHERENT],
Measurement of the Coherent Elastic Neutrino-Nucleus Scattering Cross Section on CsI by COHERENT,
\href{https://doi.org/10.1103/PhysRevLett.129.081801}{{Phys. Rev. Lett.} \textbf{129}, 081801 (2022)}.




\bibitem{Cadeddu:2019}
M. Cadeddu and F. Dordei,
Reinterpreting the weak mixing angle from atomic parity violation in view of the Cs neutron rms radius measurement from COHERENT,
\href{https://doi.org/10.1103/PhysRevD.99.033010}{Phys. Rev. D \textbf{99}, 033010 (2019)}.


\bibitem{Cadeddu:2019eta}
M.~Cadeddu, F.~Dordei, C.~Giunti, Y.~F.~Li and Y.~Y.~Zhang,
Neutrino, electroweak, and nuclear physics from COHERENT elastic neutrino-nucleus scattering with refined quenching factor,
\href{http://10.1103/PhysRevD.101.033004}{Phys. Rev. D \textbf{101}, 033004 (2020)}.



\bibitem{Khan:2023}
A. N.~Khan,
Neutrino millicharge and other electromagnetic interactions with COHERENT-2021 data,
\href{https://doi.org/10.1016/j.nuclphysb.2022.116064}{Nucl. Phys. B \textbf{986}, 116064 (2023)}.


\bibitem{Khan:2023b}
A. N.~Khan,
Light new physics and neutrino electromagnetic interactions in XENONnT,
\href{https://doi.org/10.1016/j.physletb.2022.137650}{Phys. Lett. B \textbf{837}, 137650 (2023)}.




\bibitem{Dent:2016wcr}
J.~B.~Dent, B.~Dutta, S.~Liao, J.~L.~Newstead, L.~E.~Strigari and J.~W.~Walker,
Probing light mediators at ultralow threshold energies with coherent elastic neutrino-nucleus scattering,
\href{https://doi.org/10.1103/PhysRevD.96.095007}{{Phys. Rev. D} \textbf{96}, 095007 (2017)}. 



\bibitem{AristizabalSierra:2018eqm}
D.~Aristizabal Sierra, V.~De Romeri and N.~Rojas,
COHERENT analysis of neutrino generalized interactions,
\href{https://doi.org/10.1103/PhysRevD.98.075018}{{Phys. Rev. D} \textbf{98}, 075018 (2018)}. 

\bibitem{Bischer:2019ttk}
I.~Bischer and W.~Rodejohann,
General neutrino interactions from an effective field theory perspective,
\href{https://doi.org/10.1016/j.nuclphysb.2019.114746}{{Nucl. Phys. B} \textbf{947}, 114746 (2019)}. 


\bibitem{Flores:2021kzl}
L.~J.~Flores, N.~Nath and E.~Peinado,
CE\ensuremath{\nu}NS as a probe of flavored generalized neutrino interactions,
\href{https://doi.org/10.1103/PhysRevD.105.055010}{{Phys. Rev. D} \textbf{105}, 055010 (2022)}. 


\bibitem{Barranco:2011wx}
J.~Barranco, A.~Bolanos, E.~A.~Garces, O.~G.~Miranda and T.~I.~Rashba,
Tensorial NSI and Unparticle physics in neutrino scattering,
\href{https://doi.org/10.1142/S0217751X12501473}{{Int. J. Mod. Phys. A} \textbf{27}, 1250147 (2012)}. 

\bibitem{AristizabalSierra:2017joc}
D.~Aristizabal Sierra, N.~Rojas and M.~H.~G.~Tytgat,
Neutrino non-standard interactions and dark matter searches with multi-ton scale detectors,
\href{https://doi.org/10.1007/JHEP03(2018)197}{{JHEP} \textbf{03}, 197 (2018)}. 

\bibitem{Giunti:2019xpr}
C.~Giunti,
General COHERENT constraints on neutrino nonstandard interactions,
\href{https://doi.org/10.1103/PhysRevD.101.035039}{{Phys. Rev. D} \textbf{101}, 035039 (2020)}. 

\bibitem{Mustamin:2021mtq}
M.~F.~Mustamin and M.~Demirci,
Study of Non-standard Neutrino Interactions in Future Coherent Elastic Neutrino-Nucleus Scattering Experiments,
\href{https://doi.org/10.1007/s13538-021-00867-x}{{Braz. J. Phys.} \textbf{51}, 813-819 (2021)}.


\bibitem{Harnik:2012ni}
R.~Harnik, J.~Kopp and P.~A.~N.~Machado,
Exploring nu Signals in Dark Matter Detectors,
\href{http://doi.org/10.1088/1475-7516/2012/07/026}{{JCAP} \textbf{07}, 026 (2012)}.


\bibitem{Boehm:2020ltd}
C.~B\oe{}hm, D.~G.~Cerde\~no, M.~Fairbairn, P.~A.~N.~Machado and A.~C.~Vincent,
Light new physics in XENON1T,
\href{http://10.1103/PhysRevD.102.115013}{{Phys. Rev. D} \textbf{102}, 115013 (2020)}.




\bibitem{Schwemberger:2022fjl}
T.~Schwemberger and T.~T.~Yu,
Detecting beyond the standard model interactions of solar neutrinos in low-threshold dark matter detectors,
\href{https://doi.org/10.1103/PhysRevD.106.015002}{{Phys. Rev. D} \textbf{106}, 015002 (2022)}. 


\bibitem{Coloma:2022umy}
P.~Coloma, M.~C.~Gonzalez-Garcia, M.~Maltoni, J.~P.~Pinheiro and S.~Urrea,
Constraining new physics with Borexino Phase-II spectral data,
\href{http://doi.org/10.1007/JHEP07(2022)138}{{JHEP} \textbf{07}, 138 (2022)}. 


\bibitem{Demirci:2021zci}
M.~Demirci and M.~F.~Mustamin,
Probing Light New Mediators on Coherent Elastic Neutrino-Nucleus Scattering,
\href{https://doi.org/10.31526/ACP.BSM-2021.31}{{Andromeda Proceedings}, BSM21 (2021)}.


\bibitem{Majumdar:2021vdw}A.~Majumdar, D.~K.~Papoulias and R.~Srivastava,Dark matter detectors as a novel probe for light new physics, \href{https://doi.org/10.1103/PhysRevD.106.013001}{{Phys. Rev. D} \textbf{106}, 013001 (2022)}. 


\bibitem{Farzan:2018gtr}
Y.~Farzan, M.~Lindner, W.~Rodejohann and X.~J.~Xu,
Probing neutrino coupling to a light scalar with coherent neutrino scattering,
\href{https://doi.org/10.1007/JHEP05(2018)066}{{JHEP}  \textbf{05}, 066 (2018)}.

\bibitem{Cadeddu:2020nbr}
M.~Cadeddu, N.~Cargioli, F.~Dordei, C.~Giunti, Y.~F.~Li, E.~Picciau and Y.~Y.~Zhang,
Constraints on light vector mediators through coherent elastic neutrino nucleus scattering data from COHERENT,
\href{http://doi.org/10.1007/JHEP01(2021)116}{{JHEP} \textbf{01}, 116 (2021)}.




\bibitem{AtzoriCorona:2022moj}
M.~Atzori Corona, M.~Cadeddu, N.~Cargioli, F.~Dordei, C.~Giunti, Y.~F.~Li, E.~Picciau, C.~A.~Ternes and Y.~Y.~Zhang,
Probing light mediators and $(g- 2)_{\mu}$ through detection of coherent elastic neutrino nucleus scattering at COHERENT,
\href{http://doi.org/10.1007/JHEP05(2022)109}{{JHEP} \textbf{05}, 109 (2022)}.

\bibitem{DeRomeri:2022twg}
V.~De Romeri, O.~G.~Miranda, D.~K.~Papoulias, G.~Sanchez Garcia, M.~T\'ortola and J.~W.~F.~Valle,
Physics implications of a combined analysis of COHERENT CsI and LAr data,
\href{http://doi.org/10.1007/JHEP04(2023)035}{{JHEP} \textbf{04}, 035 (2023)}.




\bibitem{Li:2022jfl}
Y.~F.~Li and S.~y.~Xia,
Constraining light mediators via detection of coherent elastic solar neutrino nucleus scattering,
\href{http://doi.org/10.1016/j.nuclphysb.2022.115737}{{Nucl. Phys. B} \textbf{977}, 115737 (2022)}. 


\bibitem{Davis:1968cp}
R.~Davis, Jr., D.~S.~Harmer and K.~C.~Hoffman,
Search for Neutrinos from the Sun,
\href{https://doi.org/10.1103/PhysRevLett.20.1205}{{Phys. Rev. Lett.} \textbf{20}, 1205-1209 (1968)}.



\bibitem{Cleveland:1998nv}
B.~T.~Cleveland, T.~Daily, R.~Davis Jr., J.~R.~Distel, K.~Lande, C.~K.~Lee, P.~S.~Wildenhain and J.~Ullman,
Measurement of the solar electron neutrino flux with the Homestake chlorine detector,
\href{http://doi.org/10.1086/305343}{{Astrophys. J.} \textbf{496}, 505-526 (1998)}.


\bibitem{SAGE:1999uje}
J.~N.~Abdurashitov \textit{et al.} [SAGE],
Measurement of the solar neutrino capture rate by SAGE and implications for neutrino oscillations in vacuum,
\href{https://doi.org/10.1103/PhysRevLett.83.4686}{{Phys. Rev. Lett.} \textbf{83}, 4686-4689 (1999)}. 


\bibitem{SNO:2002tuh}
Q.~R.~Ahmad \textit{et al.} [SNO],
Direct evidence for neutrino flavor transformation from neutral current interactions in the Sudbury Neutrino Observatory,
\href{https://doi.org/10.1103/PhysRevLett.89.011301}{{Phys. Rev. Lett.} \textbf{89}, 011301 (2002)}.



\bibitem{SNO:2008gqy}
B.~Aharmim \textit{et al.} [SNO],
An Independent Measurement of the Total Active $^8$B Solar Neutrino Flux Using an Array of $^3$He Proportional Counters at the Sudbury Neutrino Observatory,
\href{https://doi.org/10.1103/PhysRevLett.101.111301}{{Phys. Rev. Lett.} \textbf{101}, 111301  (2008)}.


\bibitem{Kamiokande-II:1989hkh}
K.~S.~Hirata \textit{et al.} [Kamiokande-II],
Observation of $^8$B Solar Neutrinos in the Kamiokande-II Detector,
\href{https://doi.org/10.1103/PhysRevLett.63.16}{{Phys. Rev. Lett.} \textbf{63}, 16 (1989)}.


\bibitem{Super-Kamiokande:2001ljr}
S.~Fukuda \textit{et al.} [Super-Kamiokande],
Solar $^8$B and hep neutrino measurements from 1258 days of Super-Kamiokande data,
\href{https://doi.org/10.1103/PhysRevLett.86.5651}{{Phys. Rev. Lett.} \textbf{86}, 5651-5655 (2001)}.


\bibitem{Borexino:2007kvk}
C.~Arpesella \textit{et al.} [Borexino],
First real time detection of $^7$Be solar neutrinos by Borexino,
\href{https://doi.org/10.1016/j.physletb.2007.09.054}{{Phys. Lett. B} \textbf{658}, 101-108 (2008)}.


\bibitem{Borexino:2008fkj}
G.~Bellini \textit{et al.} [Borexino],
Measurement of the solar $^8$B neutrino rate with a liquid scintillator target and 3 MeV energy threshold in the Borexino detector,
\href{https://doi.org/10.1103/PhysRevD.82.033006}{{Phys. Rev. D} \textbf{82}, 033006 (2010)}.


\bibitem{Bahcall:2000nu}
J.~N.~Bahcall, M.~H.~Pinsonneault and S.~Basu,
Solar models: Current epoch and time dependences, neutrinos, and helioseismological properties,
\href{https://doi.org/10.1086/321493}{{Astrophys. J.} \textbf{555}, 990-1012 (2001)}.


\bibitem{Serenelli:2016dgz}
A.~Serenelli,
Alive and well: a short review about standard solar models,
\href{https://doi.org/10.1140/epja/i2016-16078-1}{{Eur. Phys. J. A} \textbf{52}, 78 (2016)}.


\bibitem{Vinyoles:2016djt}
N.~Vinyoles, A.~M.~Serenelli, F.~L.~Villante, S.~Basu, J.~Bergstr\"om, M.~C.~Gonzalez-Garcia, M.~Maltoni, C.~Pe\~na-Garay and N.~Song,
A new Generation of Standard Solar Models,
\href{https://doi.org/10.3847/1538-4357/835/2/202}{{Astrophys. J.} \textbf{835}, 202 (2017)}.


\bibitem{Vitagliano:2019yzm}
E.~Vitagliano, I.~Tamborra and G.~Raffelt,
Grand Unified Neutrino Spectrum at Earth: Sources and Spectral Components,
\href{http://doi.org/10.1103/RevModPhys.92.045006}{{Rev. Mod. Phys.} \textbf{92}, 45006 (2020)}.


\bibitem{Bahcall:2004pz}
J.~N.~Bahcall, A.~M.~Serenelli and S.~Basu,
New solar opacities, abundances, helioseismology, and neutrino fluxes,
\href{http://doi.org/10.1086/428929}{{Astrophys. J. Lett.} \textbf{621}, L85-L88 (2005)}.


\bibitem{Bahcall:2004mq}
J.~N.~Bahcall and A.~M.~Serenelli,
How do uncertainties in the surface chemical composition of the Sun affect the predicted solar neutrino fluxes?,
\href{http://doi.org/10.1086/429883}{{Astrophys. J} \textbf{626}, 530 (2005)}.


\bibitem{Cerdeno:2016sfi}
D.~G.~Cerde\~no, M.~Fairbairn, T.~Jubb, P.~A.~N.~Machado, A.~C.~Vincent and C.~B\oe{}hm,
Physics from solar neutrinos in dark matter direct detection experiments,
\href{https://doi.org/10.1007/JHEP05(2016)118}{{JHEP} \textbf{05}, 118 (2016)} [erratum: \href{https://doi.org/10.1007/JHEP09(2016)048}{JHEP \textbf{09}, 048 (2016)}]. 


\bibitem{Goodman:1984dc}
M.~W.~Goodman and E.~Witten,
Detectability of Certain Dark Matter Candidates,
\href{https://doi.org/10.1103/PhysRevD.31.3059}{{Phys. Rev. D } \textbf{31}, 3059 (1985)}.


\bibitem{Drukier:1984vhf}
A.~Drukier and L.~Stodolsky,
Principles and Applications of a Neutral Current Detector for Neutrino Physics and Astronomy,
\href{https://doi.org/10.1103/PhysRevD.30.2295}{{Phys. Rev. D } \textbf{30}, 2295 (1984)}.


\bibitem{Drukier:1986tm}
A.~K.~Drukier, K.~Freese and D.~N.~Spergel,
Detecting Cold Dark Matter Candidates,
\href{https://doi.org/10.1103/PhysRevD.33.3495}{{Phys. Rev. D } \textbf{33}, 3495-3508 (1986)}.


\bibitem{LUX:2015abn}
D.~S.~Akerib \textit{et al.} [LUX],
Improved Limits on Scattering of Weakly Interacting Massive Particles from Reanalysis of 2013 LUX Data,
\href{http://doi.org/10.1103/PhysRevLett.116.161301}{{Phys. Rev. Lett.} \textbf{116}, 161301  (2016)}.


\bibitem{PandaX-II:2017hlx}
X.~Cui \textit{et al.} [PandaX-II],
Dark Matter Results From 54-Ton-Day Exposure of PandaX-II Experiment,
\href{http://doi.org/10.1103/PhysRevLett.119.181302}{{Phys. Rev. Lett.} \textbf{119}, 181302 (2017)}. 


\bibitem{XENON:2019gfn}
E.~Aprile \textit{et al.} [XENON],
Light Dark Matter Search with Ionization Signals in XENON1T,
\href{http://doi.org/10.1103/PhysRevLett.123.251801}{{Phys. Rev. Lett.} \textbf{123}, 251801 (2019)}.


\bibitem{XENON:2020kmp}
E.~Aprile \textit{et al.} [XENON],
Projected WIMP sensitivity of the XENONnT dark matter experiment,
\href{http://doi.org/10.1088/1475-7516/2020/11/031}{{JCAP} \textbf{11}, 031 (2020)}. 

\bibitem{LZ:2018qzl}
D.~S.~Akerib \textit{et al.} [LZ],
Projected WIMP sensitivity of the LUX-ZEPLIN dark matter experiment,
\href{http://doi.org/10.1103/PhysRevD.101.052002}{{Phys. Rev. D} \textbf{101}, 052002 (2020)}. 


\bibitem{DarkSide:2018kuk}
P.~Agnes \textit{et al.} [DarkSide],
DarkSide-50 532-day Dark Matter Search with Low-Radioactivity Argon,
\href{http://doi.org/10.1103/PhysRevD.98.102006}{{Phys. Rev. D} \textbf{98}, 102006 (2018)}. 

\bibitem{CDEX:2013kpt}
K.~J.~Kang \textit{et al.} [CDEX],
Introduction to the CDEX experiment,
\href{http://doi.org/10.1007/s11467-013-0349-1}{{Front. Phys. (Beijing)} \textbf{8}, 412-437 (2013)}. 


\bibitem{CDEX:2018lau}
H.~Jiang \textit{et al.} [CDEX],
Limits on Light Weakly Interacting Massive Particles from the First 102.8 kg ${\times}$ day Data of the CDEX-10 Experiment,
\href{http://doi.org/10.1103/PhysRevLett.120.241301}{{Phys. Rev. Lett.} \textbf{120}, 241301 (2018)}. 


\bibitem{CDEX:2019isc}
Z.~She \textit{et al.} [CDEX],
Direct Detection Constraints on Dark Photons with the CDEX-10 Experiment at the China Jinping Underground Laboratory,
\href{http://doi.org/10.1103/PhysRevLett.124.111301}{{Phys. Rev. Lett.} \textbf{124}, 111301 (2020)}. 

\bibitem{CDEX:2022mlp}
X.~P.~Geng \textit{et al.} [CDEX],
Search for exotic neutrino interactions using solar neutrinos in the CDEX-10 experiment,
\href{https://doi.org/10.1103/PhysRevD.107.112002}{{Phys. Rev. D}~\textbf{107}, 112002 (2023)}.


\bibitem{Geng:2023yei}
X.~P.~Geng, L.~T.~Yang, Q.~Yue, K.~J.~Kang, Y.~J.~Li, and H.~P.~An, \textit{et al.},
Projected sensitivity of the CDEX-50 experiment,
\href{https://doi.org/10.48550/arXiv.2309.01843}{ arXiv:2309.01843 [hep-ex]}.


\bibitem{OHare:2021utq}
C.~A.~J.~O'Hare,
New Definition of the Neutrino Floor for Direct Dark Matter Searches,
\href{https://doi.org/10.1103/PhysRevLett.127.251802}{{Phys. Rev. Lett.}     ~\textbf{127}, 251802 (2021)}.


\bibitem{Buchmuller:1991ce}
W.~Buchmuller, C.~Greub and P.~Minkowski,
Neutrino masses, neutral vector bosons and the scale of $B-L$ breaking,
\href{https://doi.org/10.1016/0370-2693(91)90952-M}{{Phys. Lett. B} \textbf{267}, 395 (1991)}.


\bibitem{Fukugita:1986hr}
M.~Fukugita and T.~Yanagida,
Baryogenesis Without Grand Unification,
\href{https://doi.org/10.1016/0370-2693(86)91126-3}{{Phys. Lett. B} \textbf{174}, 45 (1986)}.

\bibitem{Cirelli:2016rnw}
M.~Cirelli, P.~Panci, K.~Petraki, F.~Sala and M.~Taoso,
Dark Matter's secret liaisons: phenomenology of a dark U(1) sector with bound states,
\href{https://doi.org/10.1088/1475-7516/2017/05/036}{JCAP \textbf{05}, 036 (2017)}.


\bibitem{Essig:2013lka} R. Essig \textit{et al.}, Working Group Report: New Light Weakly Coupled Particles, Report No. YITP-SB-36, \href{https://doi.org/10.48550/arXiv.1311.0029}{arXiv:1311.0029 [hep-ph]}.


\bibitem{Cirelli:2013ufw}
M.~Cirelli, E.~Del Nobile and P.~Panci,
Tools for model-independent bounds in direct dark matter searches,
\href{https://doi.org/10.1088/1475-7516/2013/10/019}{{JCAP}\textbf{10}, 019 (2013)}. %

\bibitem{Abdallah:2015ter}
J.~Abdallah, H.~Araujo, A.~Arbey, A.~Ashkenazi, A.~Belyaev, J.~Berger, C.~Boehm, A.~Boveia, A.~Brennan and J.~Brooke, \textit{et al.}
Simplified Models for Dark Matter Searches at the LHC,
\href{https://doi.org/10.1016/j.dark.2015.08.001}{{Phys. Dark Univ.} \textbf{9-10}, 8-23 (2015)}. 

\bibitem{Allanach:2019} B. C. Allanach, J. Davighi and S. Melville,
An anomaly-free ATLAS: charting the space of flavour-dependent gauged U(1) extensions of the Standard Model,
\href{https://doi.org/10.1007/JHEP02(2019)082}{{JHEP} \textbf{02} (2019) 082} [erratum: \href{https://doi.org/10.1007/JHEP08(2019)064}{JHEP \textbf{08}, 064 (2019)}].


\bibitem{Mohapatra:1975} R. N. Mohapatra, J. C. Pati, Left-Right Gauge Symmetry and an Isoconjugate Model of CP Violation,
\href{https://doi.org/10.1103/PhysRevD.11.566}{{Phys. Rev. D} \textbf{11}, 566  (1975)}.

\bibitem{Mohapatra2:1975} R. N. Mohapatra, J. C. Pati, 'Natural' Left-Right Symmetry,
\href{https://doi.org/10.1103/PhysRevD.11.2558}{{Phys. Rev. D} \textbf{11}, 2558 (1975)}.


\bibitem{Davidson:1978pm}
A.~Davidson,
$B-L$ as the fourth color within an $SU(2)_L \times U(1)_R \times U(1)$ model,
\href{http://doi.org/10.1103/PhysRevD.20.776}{{Phys. Rev. D} \textbf{20}, 776 (1979)}.
%

\bibitem{Mohapatra:1980qe}
R.~N.~Mohapatra and R.~E.~Marshak,
Local B-L Symmetry of Electroweak Interactions, Majorana Neutrinos and Neutron Oscillations,
\href{http://doi.org/10.1103/PhysRevLett.44.1316}{{Phys. Rev. Lett.} \textbf{44}, 1316-1319 (1980)}
[erratum: \href{https://doi.org/10.1103/PhysRevLett.44.1644.2}{Phys. Rev. Lett. \textbf{44}, 1644(E) (1980)}].


\bibitem{Ma:1997nq}
E.~Ma,
Gauged $B - 3L_{\tau}$ and radiative neutrino masses,
\href{http://doi.org/10.1016/S0370-2693(98)00599-1}{{Phys. Lett. B } \textbf{433}, 74-81 (1998)}.


\bibitem{Ma:1998dr}
E.~Ma and U.~Sarkar,
Gauged $B - 3L_{\tau}$ and baryogenesis,
\href{https://doi.org/10.1016/S0370-2693(98)01019-3}{{Phys. Lett. B} \textbf{439}, 95-102 (1998)}.


\bibitem{Chang:2000xy}
L.~N.~Chang, O.~Lebedev, W.~Loinaz and T.~Takeuchi,
Constraints on gauged $B - 3L_{\tau}$ and related theories,
\href{https://doi.org/10.1103/PhysRevD.63.074013}{{Phys. Rev. D} \textbf{63}, 074013 (2001)}.


\bibitem{Heeck:2018nzc}
J.~Heeck, M.~Lindner, W.~Rodejohann and S.~Vogl,
Non-Standard Neutrino Interactions and Neutral Gauge Bosons,
\href{https://doi.org/10.21468/SciPostPhys.6.3.038}{{SciPost Phys.} \textbf{6}, 038 (2019)}.
%

\bibitem{Coloma:2020gfv}
P.~Coloma, M.~C.~Gonzalez-Garcia and M.~Maltoni,
Neutrino oscillation constraints on U(1)' models: from non-standard interactions to long-range forces,
\href{https://doi.org/10.1007/JHEP01(2021)114}{{JHEP} \textbf{01}, 114 (2021)} [erratum: JHEP \textbf{11}, 115 (2022)].

\bibitem{Muong-2:2006rrc}
G.~W.~Bennett \textit{et al.} [Muon g-2],
Final Report of the Muon E821 Anomalous Magnetic Moment Measurement at BNL,
\href{https://doi.org/10.1103/PhysRevD.73.072003}{{Phys. Rev. D}\textbf{73}, 072003 (2006)}.

\bibitem{CDF:2022hxs}
T.~Aaltonen \textit{et al.} [CDF],
High-precision measurement of the $W$          boson mass with the CDF II detector,
\href{https://doi.org/10.1126/science.abk1781}{Science \textbf{376}, 170 (2022)}.

\bibitem{Billard:2018jnl}
J.~Billard, J.~Johnston and B.~J.~Kavanagh,
Prospects for exploring New Physics in Coherent Elastic Neutrino-Nucleus Scattering,
\href{https://doi.org/10.1088/1475-7516/2018/11/016}{{JCAP}\textbf{11}, 016 (2018)}.


\bibitem{CONNIE:2019xid}
A.~Aguilar-Arevalo \textit{et al.} [CONNIE],
Search for light mediators in the low-energy data of the CONNIE reactor neutrino experiment,
\href{http://doi.org/10.1007/JHEP04(2020)054}{{JHEP} \textbf{04}, 054 (2020)}. 


\bibitem{CONUS:2021dwh}
H.~Bonet \textit{et al.} [CONUS],
Novel constraints on neutrino physics beyond the standard model from the CONUS experiment,
\href{http://doi.org/10.1007/JHEP05(2022)085}{{JHEP} \textbf{05}, 085 (2022)}. 


\bibitem{Bertuzzo:2017}
E. Bertuzzo, F.F. Deppisch, S. Kulkarni, Y. F. P. Gonzalez and R. Z. Funchal,  Dark matter and exotic neutrino interactions in direct detection searches,
\href{https://doi.org/10.1007/JHEP04(2017)073}{{JHEP} \textbf{04}, 073 (2017)}.



\bibitem{Han:2019zkz}
T.~Han, J.~Liao, H.~Liu and D.~Marfatia,
Nonstandard neutrino interactions at COHERENT, DUNE, T2HK and LHC,
\href{https://doi.org/10.1007/JHEP11(2019)028}{{JHEP}\textbf{11}, 028 (2019)}.



\bibitem{Bertuzzo:2022}
E. Bertuzzo, G. Grilli di Cortona, and L.M.D. Ramos, Probing light vector mediators with coherent scattering at future facilities,
\href{https://doi.org/10.1007/JHEP06(2022)075}{{JHEP} \textbf{06}, 075 (2022)}.


\bibitem{TEXONO:2009knm}
M.~Deniz \textit{et al.} [TEXONO],
Measurement of $\bar{\nu}_e$-electron scattering cross section with a CsI(Tl) scintillating crystal array at the Kuo-Sheng nuclear power reactor,
\href{http://doi.org/10.1103/PhysRevD.81.072001}{{Phys. Rev. D} \textbf{81}, 072001 (2010)}.


\bibitem{ParticleDataGroup:2022pth}
R.~L.~Workman \textit{et al.} [Particle Data Group],
Review of Particle Physics,
\href{https://doi.org/10.1093/ptep/ptac097}{{PTEP} \textbf{2022}, 083C01 (2022) }.


\bibitem{Tomalak:2021} O. Tomalak, P. Machado, V. Pandey and R. Plestid, Flavor-dependent radiative corrections in coherent elastic neutrino-nucleus scattering,
\href{https://doi.org/10.1007/JHEP02(2021)097}{{JHEP} \textbf{02}, 097 (2021)}.

\bibitem{Helm:1956zz}
R.~H.~Helm,
Inelastic and Elastic Scattering of 187-Mev Electrons from Selected Even-Even Nuclei,
\href{https://doi.org/10.1103/PhysRev.104.1466}{{Phys. Rev.} \textbf{104}, 1466-1475 (1956)}.


\bibitem{Hoferichter:2015tha}
M.~Hoferichter, J.~Ruiz de Elvira, B.~Kubis and Ulf-G. ~Mei\ss{}ner,
Matching pion-nucleon Roy-Steiner equations to chiral perturbation theory,
\href{https://doi.org/10.1103/PhysRevLett.115.192301}{{Phys. Rev. Lett.} \textbf{115}, 192301 (2015)}.



\bibitem{Langacker:2009}
P.~Langacker,
The physics of heavy $Z'$ gauge bosons,
\href{https://link.aps.org/doi/10.1103/RevModPhys.81.1199}{Rev. Mod. Phys. \textbf{81}, 1199 (2009)}.


\bibitem{Okada:2018ktp}
S.~Okada,
$Z'$ Portal Dark Matter in the Minimal $B-L$ Model,
\href{http://doi.org/10.1155/2018/5340935}{{Adv. High Energy Phys.} \textbf{2018}, 5340935 (2018)}.


\bibitem{Alves:2015pea}
A.~Alves, A.~Berlin, S.~Profumo and F.~S.~Queiroz,
Dark Matter Complementarity and the Z$^\prime$ Portal,
\href{https://doi.org/10.1103/PhysRevD.92.083004}{{Phys. Rev. D} \textbf{92}, 083004 (2015)}.

\bibitem{Allanach:2015gkd}
B.~Allanach, F.~S.~Queiroz, A.~Strumia and S.~Sun,
$Z'$ models for the LHCb and $g-2$ muon anomalies,
\href{https://doi.org/10.1103/PhysRevD.93.055045}{{Phys. Rev. D} \textbf{93} 055045, (2016)} [erratum: \href{https://doi.org/10.1103/PhysRevD.95.119902}{{Phys. Rev. D} \textbf{95} 119902, (2017)}].

\bibitem{Allanach:2023uxz}
B.~Allanach and A.~Mullin,
Plan B: new $Z'$ models for $b \rightarrow s \ell^{+} \ell^{-}$ anomalies,
\href{https://doi.org/10.1007/JHEP09(2023)173}{{JHEP} \textbf{09} 173, (2023)}.


\bibitem{deSalas:2020pgw}
P.~F.~de Salas, D.~V.~Forero, S.~Gariazzo, P.~Mart\'\i{}nez-Mirav\'e, O.~Mena, C.~A.~Ternes, M.~T\'ortola and J.~W.~F.~Valle,
2020 global reassessment of the neutrino oscillation picture,
\href{https://doi.org/10.1007/JHEP02(2021)071}{{JHEP} \textbf{02}, 071 (2021)}.


\bibitem{Collar:2021fcl}
J.~I.~Collar, A.~R.~L.~Kavner and C.~M.~Lewis,
\href{https://doi.org/10.1103/PhysRevD.103.122003}{{Phys. Rev. D} \textbf{103} 122003, (2021)}.


\bibitem{Lindhard:1963}
J. Lindhard, V. Nielsen, M. Scharff, and P. V. Thomsen, Integral Equations Governing Radiation Effects (Notes on Atomic collisions, III), \href{https://www.osti.gov/biblio/4701226}{Kgl. Danske Videnskab., Selskab. Mat. Fys. Medd. 33, 1, (1963)}.


\bibitem{Bonhomme:2022lcz}
A.~Bonhomme, \textit{et al.},
Direct measurement of the ionization quenching factor of nuclear recoils in germanium in the keV energy range,
\href{https://doi.org/10.1140/epjc/s10052-022-10768-1}{{Eur. Phys. J. C}  \textbf{82}, 815 (2022)}.



\bibitem{Essig:2018tss}
R.~Essig, M.~Sholapurkar and T.~T.~Yu,
Solar Neutrinos as a Signal and Background in Direct-Detection  Experiments Searching for Sub-GeV Dark Matter With Electron Recoils,
\href{https://doi.org/10.1103/PhysRevD.97.095029}{{Phys. Rev. D}  \textbf{97}, 095029 (2018)}.



\bibitem{Fogli:2002pt}
G.~L.~Fogli, E.~Lisi, A.~Marrone, D.~Montanino and A.~Palazzo,
Getting the most from the statistical analysis of solar neutrino oscillations,
\href{http://doi.org/10.1103/PhysRevD.66.053010}{{Phys. Rev. D} \textbf{66}, 053010 (2002)}.


\bibitem{A:2022acy}
S.~K.~A., A.~Majumdar, D.~K.~Papoulias, H.~Prajapati and R.~Srivastava,
Implications of first LZ and XENONnT results: A comparative study of neutrino properties and light mediators,
\href{http://doi.org/10.1016/j.physletb.2023.137742}{{Phys. Lett. B} \textbf{839}, 137742 (2023)}.


\bibitem{Melas:2023olz}
P.~Melas, D.~K.~Papoulias and N.~Saoulidou,
Probing generalized neutrino interactions with the DUNE Near Detector,
\href{http://doi.org/10.1007/JHEP07(2023)190}{{JHEP} \textbf{07}, 190 (2023)}.



\bibitem{Blinov:2019gcj}
N.~Blinov, K.~J.~Kelly, G.~Z.~Krnjaic and S.~D.~McDermott,
Constraining the Self-Interacting Neutrino Interpretation of the Hubble Tension,
\href{http://doi.org/10.1103/PhysRevLett.123.191102}{{Phys. Rev. Lett.} \textbf{123}, 191102  (2019)}.


\bibitem{Muong-2:2023cdq}
D.~P.~Aguillard \textit{et al.} [Muon g-2],
Measurement of the Positive Muon Anomalous Magnetic Moment to 0.20~ppm,
\href{http://doi.org/10.1103/PhysRevLett.131.161802}{{Phys. Rev. Lett.} \textbf{131}, 161802  (2023)}.


\bibitem{LSND:2001akn}
L.~B.~Auerbach \textit{et al.} [LSND],
Measurement of electron-neutrino electron elastic scattering,
\href{http://doi.org/10.1103/PhysRevD.63.112001}{{Phys. Rev. D} \textbf{63}, 112001  (2001)}.


\bibitem{CHARM-II:1994dzw}
P.~Vilain \textit{et al.} [CHARM-II],
Precision measurement of electroweak parameters from the scattering of muon-neutrinos on electrons,
\href{http://doi.org/10.1016/0370-2693(94)91421-4}{{Phys. Lett. B} \textbf{335}, 246-252  (1994)}.


\bibitem{NA64:2022yly}
Y.~M.~Andreev \textit{et al.} [NA64],
Search for a New $B-L$ $Z'$ Gauge Boson with the NA64 Experiment at CERN,
\href{http://doi.org/10.1103/PhysRevLett.129.161801}{{Phys. Rev. Lett.} \textbf{129}, 161801  (2022)}.


\bibitem{Beda:2009kx}
A.~G.~Beda \textit{et al.} [GEMMA], 
GEMMA experiment: Three years of the search for the neutrino magnetic moment,
\href{http://doi.org/10.1134/S1547477110060063}{{Phys. Part. Nucl. Lett.} \textbf{7}, 406-409 (2010)}.


\bibitem{BaBar:2014zli}
J.~P.~Lees \textit{et al.} [BaBar],
Search for a Dark Photon in $e^+e^-$ Collisions at BaBar,
\href{http://doi.org/10.1103/PhysRevLett.113.201801}{{Phys. Rev. Lett.} \textbf{113}, 201801 (2014)}.


\bibitem{LHCb:2019vmc}
R.~Aaij \textit{et al.} [LHCb], Search for $A'\to\mu^+\mu^-$ Decays,
\href{http://doi.org/10.1103/PhysRevLett.124.041801}{{Phys. Rev. Lett.} \textbf{124}, 041801 (2020)}.


\bibitem{ALICE:2012aqc}
B.~Abelev \textit{et al.} [ALICE],
Centrality Dependence of Charged Particle Production at Large Transverse Momentum in Pb-Pb Collisions at $\sqrt{s_{\rm{NN}}} = 2.76$ TeV,
\href{http://doi.org/10.1016/j.physletb.2013.01.051}{{Phys. Lett. B} \textbf{720}, 52-62 (2013)}.


\bibitem{A1:2011yso}
H.~Merkel \textit{et al.} [A1],
Search for Light Gauge Bosons of the Dark Sector at the Mainz Microtron,
\href{http://doi.org/10.1103/PhysRevLett.106.251802}{{Phys. Rev. Lett.} \textbf{106}, 251802 (2011)}.


\bibitem{PHENIX:2014duq}
A.~Adare \textit{et al.} [PHENIX],
Search for dark photons from neutral meson decays in $p + p$ and $d$ + Au collisions at $\sqrt{s_{NN}} =$ 200 GeV,
\href{http://doi.org/10.1103/PhysRevC.91.031901}{{Phys. Rev. C}  \textbf{91}, 031901 (2015)}.


\bibitem{NA482:2015wmo}
J.~R.~Batley \textit{et al.} [NA48/2],
Search for the dark photon in $\pi^0$ decays,
\href{http://doi.org/10.1016/j.physletb.2015.04.068}{{Phys. Lett. B} \textbf{746}, 178-185 (2015)}.


\bibitem{Berlin:2018bsc}
A.~Berlin, N.~Blinov, G.~Krnjaic, P.~Schuster and N.~Toro,
Dark Matter, Millicharges, Axion and Scalar Particles, Gauge Bosons, and Other New Physics with LDMX,
\href{http://doi.org/10.1103/PhysRevD.99.075001}{{Phys. Rev. D} \textbf{99}, 075001 (2019)}.


\bibitem{CCFR:1991lpl}
S.~R.~Mishra \textit{et al.} [CCFR],
Neutrino Tridents and $W-Z$ Interference,
\href{http://doi.org/10.1103/PhysRevLett.66.3117}{{Phys. Rev. Lett.} \textbf{66}, 3117-3120 (1991)}.


\bibitem{Barbieri:1975xy}
R.~Barbieri and T.~E.~O.~Ericson,
Evidence against the existence of a low mass scalar boson from neutron-nucleus scattering,
\href{http://doi.org/10.1016/0370-2693(75)90073-8}{{Phys. Lett. B} \textbf{57}, 270-272 (1975)}.


\end{thebibliography}
\end{document}